\documentclass[11pt,xcolor=dvipsnames]{article}
\pdfoutput=1 

\usepackage[colorlinks,urlcolor=black,linkcolor = blue,citecolor = black,]{hyperref}
\usepackage{subfig}
\usepackage{latexsym}
\usepackage{epsfig}
\usepackage{placeins}
\usepackage[mathscr]{eucal}
\usepackage{amsfonts}
\usepackage{amscd}
\usepackage{cite}
\usepackage{array}
\usepackage{amssymb}
\usepackage{colordvi}
\usepackage[centertags]{amsmath}
\usepackage{enumerate}
\usepackage{graphicx}
\usepackage{booktabs}
\usepackage{theorem}
\usepackage{soul}
\usepackage{mcite}
\usepackage{slashed}
\usepackage{braket}
\usepackage{units}
\usepackage{siunitx}
\usepackage{float}

\usepackage{xcolor}
\usepackage{ulem}
\usepackage{bbm}
\usepackage[utf8]{inputenc}
\usepackage{fancyvrb}
\usepackage{framed}
\usepackage{xspace}
\usepackage{todonotes}

\newcommand{\beq}{\begin{eqnarray}}
\newcommand{\eeq}{\end{eqnarray}}

\newcommand{\hc}{\text{ h.c.}}

\renewcommand{\Re}{\text{Re}\!}
\renewcommand{\Im}{\text{Im}\!}
\renewcommand{\eqref}[1]{Eq.~(\ref{#1})}

\newcommand{\cbrak}[1]{\left(#1\right)}

\allowdisplaybreaks

\usepackage[printonlyused]{acronym}

\renewcommand{\Re}{\Re\!}
\renewcommand{\Im}{\Im\!}

\newcommand{\figref}[1]{Fig.~\ref{#1}}
\renewcommand{\eqref}[1]{Eq.~(\ref{#1})}

\newcommand{\sect}[1]{Sec.~\ref{#1}}

\newcommand{\appen}[1]{App.~\ref{#1}}

\newcommand{\gev}{~\text{GeV}}

\newcommand{\s}{\newline \vspace*{-3.5mm}}

\newcommand{\nR}{\mathbb{R}}

\newcommand{\nZ}{\mathbb{Z}}

\newcommand{\ii}{\mathrm{i}}

\newenvironment{kasten*}[1]
{
\hspace{0.05\linewidth}
\begin{minipage}{0.95\linewidth}
\setlength{\fboxsep}{10pt}
\definecolor{shadecolor}{gray}{0.9}
\definecolor{framecolor}{gray}{0}

\MakeFramed {\FrameRestore}
\subsection*{#1}
}
{
\endMakeFramed
\end{minipage}
\vspace{1em}
}

\newcommand{\mhSM}{m_h}
\newcommand{\mHc}{m_{H^{\pm}}}
\newcommand{\mhI}{m_{h_{1}}}
\newcommand{\mhII}{m_{h_{2}}}

\renewcommand{\Re}{\operatorname{Re}}
\renewcommand{\Im}{\operatorname{Im}}
\newcommand{\e}{\mathrm{e}}

\usepackage[printonlyused]{acronym}

\usepackage{cleveref}
\crefname{chapter}{Chapter}{Chapter}
\crefname{section}{Sec.}{Secs.}
\crefname{table}{Tab.}{Tabs.}
\crefname{figure}{Fig.}{Figs.}
\crefname{equation}{Eq.}{Eqs.}
\crefname{appendix}{Appendix\ }{Appendix\ }

\begin{document}
\title{
\vspace*{-3.7cm}
\phantom{h} \hfill\mbox{\small KA-TP-13-2022}\\[-1.1cm]
\vspace*{2.7cm}
\textbf{Electroweak Phase Transition \\ in a Dark Sector with CP
  Violation \\[4mm]}}

\date{}
\author{
Lisa Biermann$^{1\,}$\footnote{E-mail: \texttt{lisa.biermann@kit.edu}} ,
Margarete M\"{u}hlleitner$^{1\,}$\footnote{E-mail:
\texttt{margarete.muehlleitner@kit.edu}} ,
Jonas M\"{u}ller$^{1\,}$
\\[9mm]
{\small\it
$^1$Institute for Theoretical Physics, Karlsruhe Institute of Technology,} \\
{\small\it 76128 Karlsruhe, Germany}}
\maketitle

\begin{abstract}
In this paper, we investigate the possibility of a strong first-order electroweak
phase transition (SFOEWPT) in the model {\it `CP in the Dark'}. The Higgs sector
of the model consists of two scalar doublets and one scalar singlet
with a specific discrete symmetry. After spontaneous symmetry
breaking the model has a Standard-Model-like phenomenology and a hidden scalar
sector with a viable Dark Matter candidate supplemented by explicit CP
violation that solely occurs in the hidden sector. The model `CP in the
Dark' has been 
implemented in the C++ code {\tt BSMPT~v2.3} which performs a global
minimisation of the finite-temperature one-loop corrected effective
potential and searches for SFOEWPTs. An SFOEWPT is found to be allowed
in a broad range of the parameter space. Furthermore, there are
parameter scenarios where spontaneous CP violation is generated at 
finite temperature. The in addition spontaneously broken $\nZ_2$
symmetry then leads to mixing between the dark and the visible sector
so that CP violation in the dark is promoted at finite temperature to
the visible sector and thereby provides additional sources of CP violation that are not
restricted by the electric dipole moment measurements at zero
temperature. Thus, `CP in the Dark' provides a  
promising candidate for the generation of the baryon asymmetry of the
universe through electroweak baryogenesis. 
\end{abstract}
\thispagestyle{empty}
\vfill
\newpage
\setcounter{page}{1}

\maketitle

\section{Introduction}\label{sec:Introduction}
The discovery of the Higgs boson by the LHC experiments ATLAS \cite{Aad:2012tfa}
and CMS \cite{Chatrchyan:2012xdj} structurally completes the Standard Model (SM) of
particle physics. While the discovery of the 125~GeV scalar which
behaves very SM-like marked a milestone for elementary particle
physics, the SM itself leaves a lot of open problems to be
explained. Many extensions beyond the SM, often entailing enlarged
Higgs sectors, have been proposed to explain {\it e.g.}~the existence
of Dark Matter (DM) or the matter-antimatter asymmetry of the
universe \cite{Bennett:2012zja}. The latter can be explained through the mechanism of
electroweak baryogenesis (EWBG)
\cite{Kuzmin:1985mm,Cohen:1990it,Cohen:1993nk,Quiros:1994dr,Rubakov:1996vz,Funakubo:1996dw,Trodden:1998ym,Bernreuther:2002uj,Morrissey:2012db}
provided the three Sakharov conditions \cite{Sakharov:1967dj}
are fulfilled. These require baryon number violation, C and CP
violation and departure from the thermal equilibrium. The asymmetry
can be generated if the electroweak
phase transition (EWPT), which proceeds through bubble formation, is
of strong first order \cite{Trodden:1998ym,Morrissey:2012db}. In this
case, the baryon number violating sphaleron
transitions in the false vacuum are sufficiently suppressed
\cite{Manton:1983nd,Klinkhamer:1984di}. While in the SM in
principle all Sakharov conditions are met, it can provide a
strong first-order EWPT (SFOEWPT) only for a Higgs boson mass around
70-80~GeV \cite{Kajantie:1996mn,Csikor:1998eu}. Moreover, the amount
of CP violation in the SM, originating from the
Cabibbo-Kobayashi-Maskawa (CKM) matrix in not large enough to
quantitatively reproduce the measured matter-antimatter asymmetry
\cite{Morrissey:2012db,Gavela:1993ts}. \s

In this paper, we investigate the model `CP in the Dark' with respect to
two of the above-mentioned open problems of the SM, namely Dark Matter
and the observed matter-antimatter asymmetry. The model has been
proposed for the first time in \cite{Azevedo:2018fmj}. It consists of
a 2-Higgs-Doublet Model (2HDM) \cite{Lee:1973iz,Branco:2011iw} Higgs
sector extended by a real singlet field, hence a Next-to-2HDM
(N2HDM). Compared to the N2HDM studied in \cite{Muhlleitner:2016mzt}
it uses a different discrete symmetry. The symmetry is designed such
that it leads to the following interesting properties of the model:
$(i)$ an SM-like Higgs boson $h$ that is naturally aligned because of
the vacuum of the model preserving the chosen discrete symmetry; $(ii)$ a viable DM
candidate with its stability being guaranteed by the vacuum of the model
and with its mass and couplings satisfying all existing DM search
constraints; $(iii)$ extra sources of CP violation existing solely in
the 'dark' sector.\footnote{A 3-Higgs Doublet Model with CP violation in
  the dark sector has been presented in~\cite{Cordero-Cid:2016krd,Sokolowska:2017adz}.} Because of this hidden CP violation, which is
explicit, the SM-like Higgs $h$ behaves
almost exactly like the SM Higgs boson, its couplings to massive gauge
bosons and fermions are exactly those of the SM Higgs (modulo
contributions from a high number of loops). In \cite{Azevedo:2018fmj}
we investigated the phenomenology of the model with respect to
collider and DM observables. We discussed how the hidden CP violation
might be tested at future colliders, namely through anomalous gauge
couplings. \s

This hidden CP violation can be very interesting in the context of
EWBG. First of all, with the extended Higgs sector of `CP in the Dark'
we have the chance to generate an SFOEWPT. Second, the fact that CP
violation does not occur in the visible sector at zero temperature 
means that it is not constrained by the strict bounds from the
electric dipole moment (EDM) measurements. This means in turn that CP
violation in the dark sector 
can take any value without being in conflict with experiment. If this
CP violation can be translated to the visible sector we have very
powerful means to enhance the generated baryon asymmetry of the
universe (BAU) through a large amount of CP violation. Additionally, the
model provides a viable DM candidate so that we are able to solve two
of the most prominent open questions than cannot be resolved within
the SM. \s

In this paper, we investigate the model `CP in the Dark' with respect to
its potential of generating an SFOEWPT. We furthermore analyse whether the CP
violation in the dark sector can be transmitted at finite temperature to the visible
sector and thereby improve the conditions of generating a baryon
asymmetry large enough to match the experimental value. It turns out that at
finite temperature both CP violation and violation of the discrete
symmetry are generated spontaneously through the appearance of the
corresponding non-vanishing vacuum expectation values. This allows the
transmission of  CP violation from the dark to the visible Higgs
sector. Note that in our investigation we take into account all relevant
theoretical and experimental constraints. We find in particular 
that our model not only provides an SFOEWPT with possibly spontaneous
CP violation but also complies with all 
constraints from DM observables. \s

The structure of the paper is as follows. We start by introducing the
model `CP in the Dark' in \sect{sec:Model}. In \sect{sec:SFOEWPT} the
calculation of the strength of the EWPT, as well as the chosen
renormalisation prescription is described. In \sect{sec:Numerical}  we
describe our procedure to generate viable parameter points for our
numerical analysis. We show and discuss our
results in \sect{sec:Results} and conclude in \sect{sec:Conclusion}. 

\section{CP in the Dark}\label{sec:Model}
The model `CP in the Dark' proposed in \cite{Azevedo:2018fmj} is a
specific version of the N2HDM
\cite{Chen:2013jvg,Muhlleitner:2016mzt,Engeln:2018mbg} which
features an extended scalar sector with two scalar doublets
$\Phi_1$ and $\Phi_2$ and a real scalar singlet $\Phi_S$. In contrast
to the N2HDM discussed in \cite{Muhlleitner:2016mzt},
however, the Lagrangian is required to be invariant under solely one discrete
$\nZ_2$ symmetry, namely 
\begin{align}
    \Phi_1\rightarrow\Phi_1\,,\quad\Phi_2\rightarrow-\Phi_2\,,\quad\Phi_S\rightarrow-\Phi_S\,.\label{eq:Z2Symmetry}
\end{align}
With the Yukawa sector considered to be neutral under this symmetry,
therefore only one of the two doublets, $\Phi_1$, couples to the
fermions so that the absence of scalar-mediated tree-level
flavour-changing neutral currents (FCNC) is automatically ensured, and
the Yukawa sector is identical to the SM one. With this $\nZ_2$ symmetry,
the most general scalar 
tree-level potential invariant under $\mathrm{SU}(2)_L\times
\mathrm{U}(1)_Y$ reads 
\begin{equation}
    \begin{split}
    V^{(0)}\,=\,\,\,&m_{11}^2|\Phi_1|^2+m_{22}^2|\Phi_2|^2+\frac{m_S^2}{2}\Phi_S^2+\left( A \Phi_1^\dagger \Phi_2 \Phi_S + \hc \right)\\
    &+\frac{\lambda_1}{2}|\Phi_1|^4+\frac{\lambda_2}{2}|\Phi_2|^4+\lambda_3 |\Phi_1|^2|\Phi_2|^2+\lambda_4 |\Phi_1^\dagger \Phi_2|^2+\frac{\lambda_5}{2}[(\Phi_1^\dagger \Phi_2)^2+(\Phi_2^\dagger \Phi_1)^2]\\
    &+\frac{\lambda_6}{4}\Phi_S^4+\frac{\lambda_7}{2}|\Phi_1|^2 \Phi_S^2+\frac{\lambda_8}{2}|\Phi_2|^2\Phi_S^2\,.\label{eq:TreeLevelPotential}
    \end{split}
\end{equation}
All parameters of the potential are real, except for the trilinear
coupling $A$. A possible complex phase of the quartic coupling
$\lambda_5$ is absorbed by a proper rotation of the doublet fields
\cite{Azevedo:2018fmj}. After electroweak symmetry breaking (EWSB), the
scalar doublets and the singlet can be expanded around the vacuum
expectation values (VEVs). Allowing for the most
general vacuum structure where besides the doublet and singlet VEVs
$\overline{\omega}_{1,2}$ and $\overline{\omega}_S$, respectively, we take into account also
CP-violating ($\overline{\omega}_{\text{CP}}$) and charge-breaking 
($\overline{\omega}_{\text{CB}}$) VEVs, we have 
\begin{align}
    \Phi_1=\frac{1}{\sqrt{2}}\begin{pmatrix}\rho_1+i\eta_1\\\zeta_1+\overline{\omega}_1+i\Psi_1\end{pmatrix}\,,\quad\Phi_2=\frac{1}{\sqrt{2}}\begin{pmatrix}\rho_2+\overline{\omega}_\text{CB}+i\eta_2\\\zeta_2+ \overline{\omega}_2+i(\Psi_2+\overline{\omega}_\text{CP})\end{pmatrix}\,,\quad\Phi_S=\zeta_S+\overline{\omega}_\text{S}\,.\label{eq:VacuumTFinite}
\end{align}
Here we introduced the charged CP-even and CP-odd fields $\rho_i$
and $\eta_i$ as well as the neutral CP-even and CP-odd fields
$\zeta_i$ and $\Psi_i$ ($i=1,2$).
The zero-temperature vacuum structure is chosen such that the imposed
$\nZ_2$-symmetry remains unbroken, {\it i.e.}
\begin{align}
    \langle
  \Phi_1\rangle|_{T=\SI{0}{GeV}}=\frac{1}{\sqrt{2}}\begin{pmatrix}0\\v_1\end{pmatrix},\quad
  \langle
  \Phi_2\rangle|_{T=\SI{0}{GeV}}=\begin{pmatrix}0\\0\end{pmatrix}\quad\text{and}\quad
  \langle \Phi_S\rangle|_{T=\SI{0}{GeV}}=0\,,\label{eq:VacuumTZero} 
\end{align}
with $\overline{\omega}_1 |_{T=\SI{0}{GeV}}\equiv v_1 \equiv v\approx\SI{246.22}{GeV}$. Since
only the first Higgs doublet couples to fermions, $\Phi_1$ is the
SM-like doublet and provides an SM-like Higgs boson $h$. The first
doublet can be written in terms of the mass eigenstates as 
\begin{align}
    \Phi_1=\begin{pmatrix}G^+\\\frac{1}{\sqrt{2}}\left(h+v_1+iG^0\right)\end{pmatrix}\,,
\label{eq:MassEigenstates}
\end{align}
where $G^+$ denotes the massless charged and $G^0$ the massless
neutral Goldstone boson, respectively. 
The upper charged components of the second doublet $\Phi_2$ provide
the charged mass eigenstate $H^+$.  
The neutral fields $\zeta_2$, $\Psi_2$ and $\zeta_S$ mix,  
with the corresponding mixing matrix given by
\begin{align}
    M_N^2=\begin{pmatrix}
        m_{22}^2+\frac{v_1^2}{2}\lambda_{345}&0&\Re\, (A)v_1\\
        0&m_{22}^2+\frac{v_1^2}{2}\overline{\lambda}_{345}&-\Im\, (A)v_1\\
        \Re\, (A)v_1&-\Im\, (A)v_1&m_S^2+\frac{v_1^2}{2}\lambda_7 \\
    \end{pmatrix}\,,
\end{align}
where we have introduced $\lambda_{345} \equiv \lambda_3 + \lambda_4 + \lambda_5$ and $\overline{\lambda}_{345} \equiv \lambda_3 + \lambda_4 -
\lambda_5$. Diagonalisation with the rotation matrix $R$ yields the
mass eigenvalues $h_i$ ($i=1,2,3$),
\begin{align}
    \text{diag}(m_{h_1}^2,m_{h_2}^2,m_{h_3}^2) =  R M_N^2 R^T \,,
\end{align} 
which by convention are ordered by ascending mass, $m_{h_1}\leq m_{h_2}\leq m_{h_3}$.
The orthogonal matrix $R$ can be parametrised in terms of three mixing angles
$\alpha_i$ ($i=1,2,3$),
\begin{align}
    R = \begin{pmatrix}
        c_{\alpha_1} c_{\alpha_2} & s_{\alpha_1} c_{\alpha_2} & s_{\alpha_2}\\
        -(c_{\alpha_1}s_{\alpha_2}s_{\alpha_3}+s_{\alpha_1}c_{\alpha_3})&c_{\alpha_1}c_{\alpha_3}-s_{\alpha_1}s_{\alpha_2}s_{\alpha_3}&c_{\alpha_2}s_{\alpha_3}\\
        -c_{\alpha_1}s_{\alpha_2}c_{\alpha_3}+s_{\alpha_1}s_{\alpha_3}&-(c_{\alpha_1}s_{\alpha_3}+s_{\alpha_1}s_{\alpha_2}c_{\alpha_3})&c_{\alpha_2}c_{\alpha_3}
    \end{pmatrix}\,.\label{eq:RotationMatrix}
\end{align}
Note that the chosen vacuum at zero temperature  given in
\eqref{eq:VacuumTZero} does not include a CP-violating VEV.
As mentioned above, the zero-temperature vacuum also  
conserves the $\nZ_2$ symmetry of \eqref{eq:Z2Symmetry}, introducing a conserved quantum number, called \textit{dark charge}. 
While all SM-like particles have dark charge $+1$, the charged scalar
$H^+$ and the neutral scalars $h_{1,2,3}$ that originate from the
second doublet $\Phi_2$, and the real singlet $\Phi_S$, have dark
charge $-1$. They are called \textit{dark particles}.  
The lightest neutral dark particle $h_1$ therefore acts as a
\textit{stable} dark matter candidate. 
`CP in the Dark' additionally features \textit{explicit} CP violation in
the dark sector which is introduced through $\Im A \neq 0$. For
further details, we refer to \cite{Azevedo:2018fmj}. \s

The model `CP in the Dark' is determined by thirteen input parameters. 
Exploiting the minimum condition
\begin{eqnarray}
  m_{11}^2+\frac{1}{2}\lambda_1 v_1^2 = 0
\end{eqnarray}
to trade $m_{11}^2$ for $v_1$, we use the following input parameters
for the parameter scan performed with {\tt ScannerS} \cite{Coimbra:2013qq,Ferreira:2014dya,Muhlleitner:2016mzt,Muhlleitner:2020wwk}, {\it
  cf.}~Sec.~\ref{sec:Numerical}, 
\begin{eqnarray}
m_h \,, \ m_{h_1} \,, \ m_{h_2} \,, \ m_{H^\pm} \,, \ v_1 \,, \
  m_{22}^2 \,, \ m_S^2 \,,\ \alpha_1 \,, \ \alpha_2 \,,\ \alpha_3 \,,\
  \lambda_2 \,,\ \lambda_6 \;, \ \lambda_8  \;.
\end{eqnarray}
For our program {\tt BSMPT}~\cite{Basler:2018cwe,Basler:2020nrq} used
to compute the EWPT of the model, we use the default input
set of {\tt BSMPT},
\begin{eqnarray}
m_h \,, \ m_S^2 \,, \ m_{22}^2 \,, \ v_1 \,, \ \Re A \,, \ \Im A \,,
\ \lambda_2 \,, \ \lambda_3 \,, \ \lambda_4 \,, \ \lambda_5 \,, \ \lambda_6 \,, \ \lambda_7 \,, \ \lambda_8 \;.
\end{eqnarray}

\section{Calculation of the Strength of the Phase Transition}\label{sec:SFOEWPT}
We follow the approach of
Refs.~\cite{Basler:2016obg,Basler:2017uxn,Basler:2019iuu} to determine the
strength $\xi_c$ of the EWPT. It is given by the ratio of the critical VEV
$v_c$ at the critical temperature $T_c$, $\xi_c=v_c/T_c$. 
The critical temperature is
defined as the temperature where the symmetric and the broken minimum
are degenerate, hence  
\begin{align}
    V(v=0,T_c) = V(v=v_c,T_c)\,,
\end{align}
with the critical VEV determined as $v_c\equiv v(T)\big|_{T=T_c}$, see
below. 
A value larger than one is indicative for a strong first-order EWPT 
\cite{Quiros:1994dr,Moore:1998swa}.\footnote{For discussions on the
  gauge dependence of $\xi_c$, we refer to
  Refs.~\cite{Dolan:1973qd,gauge2,gauge3,gauge4}.}
For the determination of $\xi_c$ we use the {\tt C++} program {\tt
  BSMPT} \cite{Basler:2018cwe,Basler:2020nrq} where we implemented the 
daisy-corrected one-loop effective potential at finite temperature of
`CP in the Dark'. `CP in the Dark' was implemented as a new model class in
{\tt BSMPT}, which is publicly available since version 2.3. Since the
corresponding details of the calculation do not differ compared to the
other models, we refer to our previous publications for further
details, {\it cf.}~\cite{Basler:2018cwe,Basler:2020nrq}. However, note
that we had to adapt the proposed 
renormalisation scheme applied in the previous publications, which we
will discuss in the following.  \s

For an efficient parameter scan, it is convenient to renormalise the
loop-corrected masses and angles to their tree-level values. 
This is usually achieved by adapting the renormalisation scheme that
has been introduced for the 2HDM in Ref.~\cite{Basler:2016obg}.  
The scheme was further applied to the C2HDM and the N2HDM in \cite{Basler:2017uxn,Basler:2018cwe,Basler:2019iuu,Basler:2021kgq} and to the CxSM in \cite{Basler:2020nrq}.
In the following, we shortly summarise the procedure and adapt it to the model `CP in the Dark'. The one-loop corrected effective potential with one-loop masses and angles renormalised to their tree-level values is constructed as
\begin{align}
    V(\omega, T) = V^{(0)}(\omega) + V^\text{CW}(\omega) + V^\text{T}(\omega, T) + V^\text{CT}(\omega)\,.\label{eq:1LEPACT}
\end{align}
The tree-level potential $V^{(0)}$ is given in
Eq.~(\ref{eq:TreeLevelPotential}) with the doublet and singlet fields
replaced by the classical constant field configurations
$\Phi_1^c=(0,\omega_1/\sqrt{2})^T$,
$\Phi_2^c=1/\sqrt{2}\,(\omega_{\text{CB}},\omega_2 +
i\omega_{\text{CP}})^T$ and $\Phi_S^c=\omega_S$, respectively.  The Coleman-Weinberg
potential is denoted by $V^\text{CW}(\omega)$, the contribution $V^\text{T}(\omega,
T)$ accounts for the thermal corrections at finite temperature, and
the counterterm potential $V^\text{CT}(\omega)$ is given by
\cite{Basler:2016obg} 
\begin{align}
    V^\text{CT} = \sum_{i=1}^{N_p}\frac{\partial V^{(0)}}{\partial p_i}\delta p_i+\sum_{k=1}^{N_v}\delta T_k (\phi_k + \omega_k)\,. \label{eq:VCTgeneralConv}
\end{align}
The number of parameters $p_i$ of the tree-level potential is denoted
by $N_p$. The finite counterterm parameters are referred to as $\delta
p_i$. Additionally, a tadpole counterterm $\delta T_k$ for each of the
$N_v=5$ field directions that are allowed to develop a non-zero VEV is
included. Applying \eqref{eq:VCTgeneralConv} to the tree-level
potential of `CP in the Dark' of \eqref{eq:TreeLevelPotential} results
in  
\begin{eqnarray}
        V^{\text{CT}}&=&\delta m_{11}^2|\Phi_1|^2+\delta
                         m_{22}^2|\Phi_2|^2+\frac{\delta
                         m_S^2}{2}\Phi_S^2 \nonumber \\
        &+&[\delta \Re\, (A)+ i\delta \Im\, (A)]\Phi_1^\dagger \Phi_2 \Phi_S + [\delta \Re\, (A)- i\delta \Im\, (A)] \Phi_S \Phi_2^\dagger \Phi_1 \nonumber \\
        &+&\frac{\delta \lambda_1}{2}|\Phi_1|^4+\frac{\delta \lambda_2}{2}|\Phi_2|^4+\delta \lambda_3 |\Phi_1|^2|\Phi_2|^2+\delta \lambda_4 |\Phi_1^\dagger \Phi_2|^2+\frac{\delta \lambda_5}{2}[(\Phi_1^\dagger \Phi_2)^2+(\Phi_2^\dagger \Phi_1)^2]\nonumber \\
        &+&\frac{\delta \lambda_6}{4}\Phi_S^4+\frac{\delta \lambda_7}{2}|\Phi_1|^2 \Phi_S^2+\frac{\delta \lambda_8}{2}|\Phi_2|^2\Phi_S^2 \nonumber \\
        &+&\delta T_\text{CB} (\rho_2+\omega_\text{CB})+\delta T_1 (\zeta_1+\omega_1) \nonumber \\
        &+&\delta T_2(\zeta_2 + \omega_2)+\delta T_\text{CP}(\Psi_2 + \omega_\text{CP})+\delta T_\text{S}(\zeta_S+\omega_\text{S})\,.\label{eq:VCTconventional}
\end{eqnarray}
The counterterm parameters $\delta p_i$ and $\delta T_k$ are
determined through the renormalisation conditions  
\begin{subequations}
    \label{eq:RenCondCWCT}
    \begin{align}
        \partial_{\phi_i}\left(V^{\text{CT}}+V^{\text{CW}}\right)\big|_{\phi=\langle \phi^c \rangle|_{T=\SI{0}{GeV}}}=0\label{nabla}\\
        \partial_{\phi_i}\partial_{\phi_j}\left(V^{\text{CT}}+V^{\text{CW}}\right)\big|_{\phi=\langle \phi^c \rangle|_{T=\SI{0}{GeV}}}=0\,.\label{hesse}
    \end{align}
\end{subequations}
In the following, we use the notation $N^{\text{CW}}_\phi \equiv \partial_\phi V^{\text{CW}}$ and $H^{\text{CW}}_{\phi_i \phi_j} \equiv \partial_{\phi_i \phi_j} V^{\text{CW}}$.
The scalar fields $\phi_i$ in the gauge basis are labelled as
\begin{align}
    \phi_i = \{\rho_1,\eta_1,\rho_2,\eta_2,\zeta_1,\Psi_1,\zeta_2,\Psi_2,\zeta_S\}\,.
\end{align}
The field configuration at $T=0$~GeV is denoted by $\langle
\phi^c\rangle|_{T=\SI{0}{GeV}}$ and given by
\begin{align}
    \langle \phi^c\rangle|_{T=\SI{0}{GeV}}=\{0,0,0,0,v_1,0,0,0,0\}\quad\text{and}\quad v_1\equiv v \approx \SI{246.22}{GeV}\,.
\end{align}
The renormalisation conditions of \eqref{eq:RenCondCWCT} yield an
overconstrained system of equations. Its five-dimensional solution
space can be parametrised by $\delta\lambda_{i}\equiv t_{i}\in\nR$
($i=2,3,6,7,8$) so that we have
\begin{subequations}
    \label{eq:CTConvScheme}
    \begin{align}
        \delta m_{11}^2 &= \frac{1}{2}H^\text{CW}_{\zeta_1\zeta_1}-\frac{3}{2}H^\text{CW}_{\rho_1\rho_1}\label{eqCTm11st}\\
        \delta m_{22}^2 &= -H^\text{CW}_{\rho_2\rho_2}-\frac{1}{2}t_{3} v_1^2\label{eqCTm22st}\\
        \delta m_S^2 &= -H^\text{CW}_{\zeta_S\zeta_S}-\frac{1}{2}t_{7}v_1^2\label{eqCTmSst}\\
        \delta \Re\, (A) &= -\frac{1}{v_1}H^\text{CW}_{\zeta_2\zeta_S}\label{eqCTReAt}\\
        \delta \Im\, (A) &= \frac{1}{v_1}H^\text{CW}_{\Psi_2\zeta_S}\label{eqCTImAt}\\
        \delta \lambda_1 &= \frac{1}{v_1^2}\left(-H^\text{CW}_{\zeta_1\zeta_1}+H^\text{CW}_{\rho_1\rho_1}\right)\label{eqCTL1t}\\
        \delta \lambda_2 &= t_{2}\label{eqCTL2t}\\
        \delta \lambda_3 &= t_{3}\label{eqCTL3t}\\
        \delta \lambda_4 &= \frac{1}{v_1^2}\left(2H^\text{CW}_{\rho_2\rho_2}-H^\text{CW}_{\zeta_2\zeta_2}-H^\text{CW}_{\Psi_2\Psi_2}\right)\label{eqCTL4t}\\
        \delta \lambda_5 &= \frac{1}{v_1^2}\left(-H^\text{CW}_{\zeta_2\zeta_2}+H^\text{CW}_{\Psi_2\Psi_2}\right)\label{eqCTL5t}\\
        \delta \lambda_6 &= t_{6}\label{eqCTL6t}\\
        \delta \lambda_7 &= t_{7}\label{eqCTL7t}\\
        \delta \lambda_8 &= t_{8}\label{eqCTL8t}\\
        \delta T_\text{CB} &= -N^\text{CW}_{\rho_2}\label{eqCTTCBt}\\
        \delta T_1 &= v_1H^\text{CW}_{\rho_1\rho_1}-N^\text{CW}_{\zeta_1}\label{eqCTT1t}\\
        \delta T_2 &= -N^\text{CW}_{\zeta_2}\label{eqCTT2t}\\
        \delta T_\text{CP} &= -N^\text{CW}_{\Psi_2}\label{eqCTTCPt}\\
        \delta T_\text{S} &=-N^\text{CW}_{\zeta_S}\,.\label{eqCTTSt}
    \end{align}
\end{subequations}
Equation~(\ref{eq:CTConvScheme}) provides a consistent solution to
\eqref{eq:RenCondCWCT} only if additionally the following identities
are fulfilled, 
\begin{subequations}
    \label{eq:HCWIdentities}
    \begin{align}
        H^\text{CW}_{\rho_1\rho_1} &= H^\text{CW}_{\Psi_1\Psi_1}\\
        H^\text{CW}_{\eta_1\eta_1} &= H^\text{CW}_{\Psi_1\Psi_1}\\
        H^\text{CW}_{\rho_2\rho_2} &= H^\text{CW}_{\eta_2\eta_2}.
    \end{align}
\end{subequations}
However, with $V^\text{CT}$ determined through \eqref{eq:CTConvScheme} we still observe 
\begin{align}
    \partial_{\zeta_2} \partial_{\Psi_2}
  \left(V^\text{CW}+V^\text{CT}\right)\big|_{\phi=\langle \phi^c
  \rangle|_{T=\SI{0}{GeV}}} \sim \mathcal{O}(10^1 \mbox{ GeV}^2)\,.\label{eq:ProblCoupling}
\end{align}
These non-cancelled second derivatives are due to field directions
$ij=\{\zeta_2 \Psi_2\}$ in which the Coleman-Weinberg potential yields
non-zero contributions $\partial_{ij}^2 V^\text{CW}$, but the
counterterm potential with the ansatz of Eq.~(\ref{eq:VCTconventional}) vanishes, $\partial^2_{ij} V^\text{CT}=0$. Hence, there is no cancellation possible. Therefore, the solution of \eqref{eq:CTConvScheme} is insufficient for these $ij$. 
The reason are loop-induced CP-violating couplings that lead to
non-zero derivatives of $V^\text{CW}$ in field directions that are not
present in the chosen $V^\text{CT}$ of Eq.~(\ref{eq:VCTconventional}). 
For `CP in the Dark' we therefore propose a \textit{modified} renormalisation
scheme by adding one additional counterterm that 
parametrises the additional CP-violating structure 
\begin{align}
    \widetilde{V}^\text{CT} \equiv V^\text{CT} + \frac{i}{2} \delta\Im(\lambda_5)\left((\Phi_1^\dagger \Phi_2)^2 - (\Phi_1 \Phi_2^\dagger)^2\right)\,. \label{eq:CTImL5}
\end{align}
The new counterterm $\delta\Im(\lambda_5)$ is constrained through \eqref{eq:RenCondCWCT} to be
\begin{align}
    \delta \Im\,(\lambda_5) = \frac{2}{v_1^2} H^\text{CW}_{\zeta_2
  \Psi_2} \,.\label{eq:CTImL5determined}
\end{align}
Furthermore, we choose the free parameters for the solution to be
$t_i=0$ for the analysis in 
\sect{sec:Numerical}. This choice corresponds to the renormalisation
scheme with all additional finite pieces set to zero that are
irrelevant for the initial aim of fixing next-to-leading order (NLO) masses and angles at
their tree-level values. \s

By performing a global minimisation of the renormalised one-loop
corrected effective potential, {\tt BSMPT} calculates the critical
temperature $T_c$ and the critical VEV $v_c$. For details, we refer to
\cite{Basler:2018cwe,Basler:2020nrq}. 
The temperature-dependent electroweak VEV $v(T)$ is
  calculated taking into account all possible $SU(2)_L$ VEV contributions,
\begin{align}
    v(T) =
  \sqrt{\overline{\omega}_1^2(T)+\overline{\omega}_2^2(T)+\overline{\omega}_{\text{CP}}^2(T)+
\overline{\omega}_{\text{CB}}^2(T)}\,.\label{eq:EWVEV}
\end{align}
Remind that the $\overline{\omega}_i$ are the field configurations that
minimise the loop-corrected effective potential at non-zero
temperature. We do not include the singlet VEV $\bar{\omega}_S$ in
Eq.~(\ref{eq:EWVEV}), but we take it into account for the minimisation
procedure. The reason is that the electroweak sphaleron couples only
to particles charged under $SU(2)_L$. 
In case the baryon-wash-out condition is fulfilled, {\it i.e.}
\begin{align}
    \xi_c = \frac{v_c}{T_c} \gtrsim 1\,,
\end{align}
the EWPT is an SFOEWPT and provides the necessary departure from
thermal equilibrium as required by the Sakharov conditions.

\section{Numerical Analysis}\label{sec:Numerical}
For our numerical analysis we perform a scan in the parameter space of
the model and keep only those points that are compatible with the
relevant theoretical and experimental constraints. We require the
SM-like Higgs boson $h$ to have a mass of $m_h=125.09\gev$
\cite{HiggsAtlas} and behave SM-like.  The remaining input parameters
of `CP in the Dark' are varied in the ranges given in Tab.~\ref{tab:input}. 
\begin{table}[b]
	\centering
	\begin{tabular}{c c c c c c}
		\toprule
		$\mhSM$  & $\mhI$                & $\mhII$                 & $\mHc$                 & $m_{22}^2$             & $m_{S}^2$ \\
		in $\gev$& in $\gev$             &  in $\gev$              & in $\gev$              & in $\gev^2$            &  in $\gev^2$ \\\midrule
		$125.09$ & $\left[1,1000\right]$ & $\left[1,1000\right]$  & $\left[65,1000\right]$  & $\left[0,10^6\right]$  & $\left[0,10^6\right]$ \\\bottomrule
		$\alpha_1$  & $\alpha_2$ & $\alpha_3$ & $\lambda_2$ & $\lambda_6$ & $\lambda_8$\\\midrule
		$\left[-\frac{\pi}{2} , \frac{\pi}{2}\right)$ &  $\left[-\frac{\pi}{2} , \frac{\pi}{2}\right)$ &  $\left[-\frac{\pi}{2} , \frac{\pi}{2}\right)$ & $\left[0,9\right]$ & $\left[0,17\right]$ & $\left[-26,26\right]$\\\bottomrule
	\end{tabular}
	\caption{Parameter scan ranges of the `CP in the Dark' input parameters 
		used in {\tt ScannerS}.}
	\label{tab:input}
	\vspace*{0.5cm}
\end{table}
For the SM parameters, we use the fine structure constant at the scale
of the $Z$ boson mass \cite{Agashe:2014kda,LHCHXSWG},
\begin{equation}
	\alpha_{\text{EM}}^{-1}(M_Z^2) = 128.962 \;,
\end{equation}
and the masses for the massive gauge bosons are chosen as \cite{Agashe:2014kda,LHCHXSWG}
\begin{equation}
	m_W = 80.385\gev \quad \text{and }\quad m_Z=91.1876\gev\,.
\end{equation}
The lepton masses are set to \cite{Agashe:2014kda,LHCHXSWG}
\begin{equation}
	m_e=0.511~\mathrm{MeV},\quad m_{\mu}=105.658~\mathrm{MeV},\quad m_{\tau} = 1.777\gev\,,
\end{equation}
and the light quark masses to \cite{LHCHXSWG}
\begin{equation}
	m_u = m_d = m_s = 100~\mathrm{MeV}\,.
\end{equation}
To be consistent with the CMS and ATLAS analyses, we take the on-shell top quark mass as \cite{LHCHXSWG,Dittmaier:2011ti}
\begin{equation}
	m_t = 172.5\gev\,
\end{equation}
and the recommended charm and bottom quark on-shell masses \cite{LHCHXSWG}
\begin{equation}
	m_c=1.51\gev \quad \text{and} \quad m_b = 4.92\gev\,.
\end{equation}
We choose the complex parametrisation of the CKM matrix
\cite{Chau:1984fp,Agashe:2014kda}, 
\begin{equation}
    V_{\text{CKM}} =
\begin{pmatrix}
    c_{12}c_{13} & s_{12}c_{13} & s_{13}\e^{-\ii \delta}\\
    -s_{12}c_{23}-c_{12}s_{23}s_{13}\e^{\ii\delta} & c_{12}c_{23}-s_{12}s_{23}s_{13} \e^{\ii \delta} & s_{23}c_{13}\\
    s_{12}s_{23}-c_{12}c_{23}s_{13}\e^{\ii\delta} & - c_{12}s_{23}-s_{12}c_{23}c_{13}\e^{\ii\delta} & c_{23}c_{13}
\end{pmatrix}\,,
  \end{equation}
where $s_{ij}\equiv\sin\theta_{ij}$ and
$c_{ij}\equiv\cos\theta_{ij}$. The angles are given in terms of the
Wolfenstein parameters as
\begin{align}
	s_{12} = \lambda \,,\quad && s_{13}\e^{\ii \delta} = A \lambda^3\cbrak{\rho+\ii \eta} \,,\quad &&s_{23} = A \lambda^2 \,,
\end{align}
with \cite{Basler:2019iuu}
\begin{align}
	\lambda =0.22537\,,&& A =0.814 \;,&& \rho = 0.117 \;,&& \eta =  0.353\,.
\end{align}
Note that we take into account a complex phase $\delta$ in the CKM matrix as an
additional source for CP violation. The impact of the
complex CKM phase compared to that of the complex phase induced by the
VEV configuration is negligible, however. 
Finally, the electroweak VEV is set to
\begin{equation}
	v_1 \equiv v=1/\sqrt{\sqrt{2}G_F} \approx 246.22\gev\,.
\end{equation}

The parameter points under investigation have to fulfil experimental
and theoretical constraints. For the generation of such parameter
points we use the {\tt C++} program {\tt ScannerS}
\cite{Coimbra:2013qq,Ferreira:2014dya,Muhlleitner:2016mzt,Muhlleitner:2020wwk}. 
{\tt ScannerS} allows us to check for boundedness from below of the tree-level
potential and uses the tree-level discriminant of
\cite{Ivanov:2015nea} to ensure the electroweak vacuum to be the
global minimum at tree level. It also checks for perturbative unitarity.
By using \texttt{BSMPT} it is also
possible to check for the NLO electroweak vacuum to be the global
minimum of the potential. Only parameter points providing a stable NLO
electroweak vacuum at zero temperature are taken into account for the
analysis. 
The check for consistency with recent flavour constraints is obsolete
since in `CP in the Dark' the charged Higgs boson belongs to the dark
sector and does not couple to fermions so that all $B$-physics
constraints are fulfilled per default. 
The compatibility with the Higgs measurements is taken into account by
{\tt ScannerS} through {\tt
  HiggsBounds}~\cite{Bechtle:2008jh,Bechtle:2011sb,Bechtle:2012lvg,Bechtle:2013wla,Bechtle:2015pma,Bechtle:2020pkv}
and {\tt HiggsSignals}~\cite{Bechtle:2013xfa,Bechtle:2014ewa}. For the
parameter scan 
the versions {\tt HiggsBounds5.9.0} and {\tt HiggsSignals2.6.1} are
used. 
We have also taken into account the latest CMS results \cite{CMS:2021kom} on the
Higgs signal strength in the photonic final state that were not
implemented in {\tt HiggsSignals}. 
We have checked the compatibility with the recent ATLAS result on the Higgs decay into
invisible particles, 
$\text{BR}(h\rightarrow\,\text{inv.})<\num{0.11}$~\cite{ATLAS:2020kdi}.
Further Dark Matter observables like the
spin-independent DM-nucleon cross section and the relic abundance are
checked with {\tt MicrOMEGAs5.2.7a}
\cite{Belanger:2006is,Belanger:2008sj,
  Belanger:2010gh,Belanger:2010pz,Belanger:2013oya,Belanger:2014vza,Barducci:2016pcb,Belanger:2018ccd,Belanger:2020gnr,MicrOMEGAsWebpage}. 
We demand that our parameter scenarios do not lead to relic densities
above the experimentally measured value of \cite{Aghanim:2018eyx}
\begin{eqnarray}
\Omega_\text{obs} h^2 = 0.1200\pm0.0012 \,,
\end{eqnarray}
while they may be below the value assuming other DM particles
beyond our model to saturate the relic density. 
Note that we need not check for the strict constraints 
on CP violation arising from the electric dipole moment measurements, where
the stringest one is provided by the ACME collaboration~\cite{Andreev:2018ayy},
as in our model CP violation beyond the SM arises only in the dark sector.
For the determination of the strength of 
the EWPT we use the new model implementation in {\tt BSMPT~v2.3}
\cite{Basler:2018cwe,Basler:2020nrq}. The code can be
  downloaded from the url:
\begin{center}
  \url{https://github.com/phbasler/BSMPT}.
\end{center}

\section{Results}\label{sec:Results}
In this section, we present our numerical results. We start with the
discussion of the visible sector of `CP in the Dark', namely in
\cref{subsec:BRhSM} with the discussion of the 
impact of the applied constraints on the branching ratios and signal
strengths of the SM-like Higgs boson in the context of the additional
dark sector. We further discuss the impact of the strength of the
phase transition. Subsequently, we investigate in \cref{subsec:SFOEWPTresults} the mass
distributions of our dark sector particles and if there is an
interplay between their masses and the requirement of an SFOEWPT. We
then analyse in detail in \cref{subsec:VEVEVOresults} the VEV
configurations that were found to minimise the 
one-loop corrected effective potential at finite temperature. Special
attention is paid here on the spontaneous generation of CP violation. 
In \cref{subsec:DMobservables} finally we show results for the DM observables and
study their interplay with the requirement of an SFOEWPT. 

\FloatBarrier
\subsection{Branching Ratios and Signal Strengths of the SM-like Higgs
  Boson}\label{subsec:BRhSM}
Since the tree-level couplings of the SM-like Higgs $h$ in our model are identical to
those of the SM Higgs boson, it is only the presence of the dark
particles that can change the branching ratios of $h$. Thus, the dark
charged Higgs boson can contribute to the decay width into a photon
pair, {\it cf.}~\cite{Azevedo:2018fmj}, so that the decay width
$\Gamma(h\to \gamma\gamma)$ is changed.\footnote{The total width
is barely changed, as the partial decay width into photons is
very small.} 
Furthermore, $h$ can decay into a pair of DM particles if 
kinematically allowed which would change the total width and hence
also the branching ratio. \s

\begin{figure}[h!]
	\centering
\includegraphics[width=0.46\textwidth]{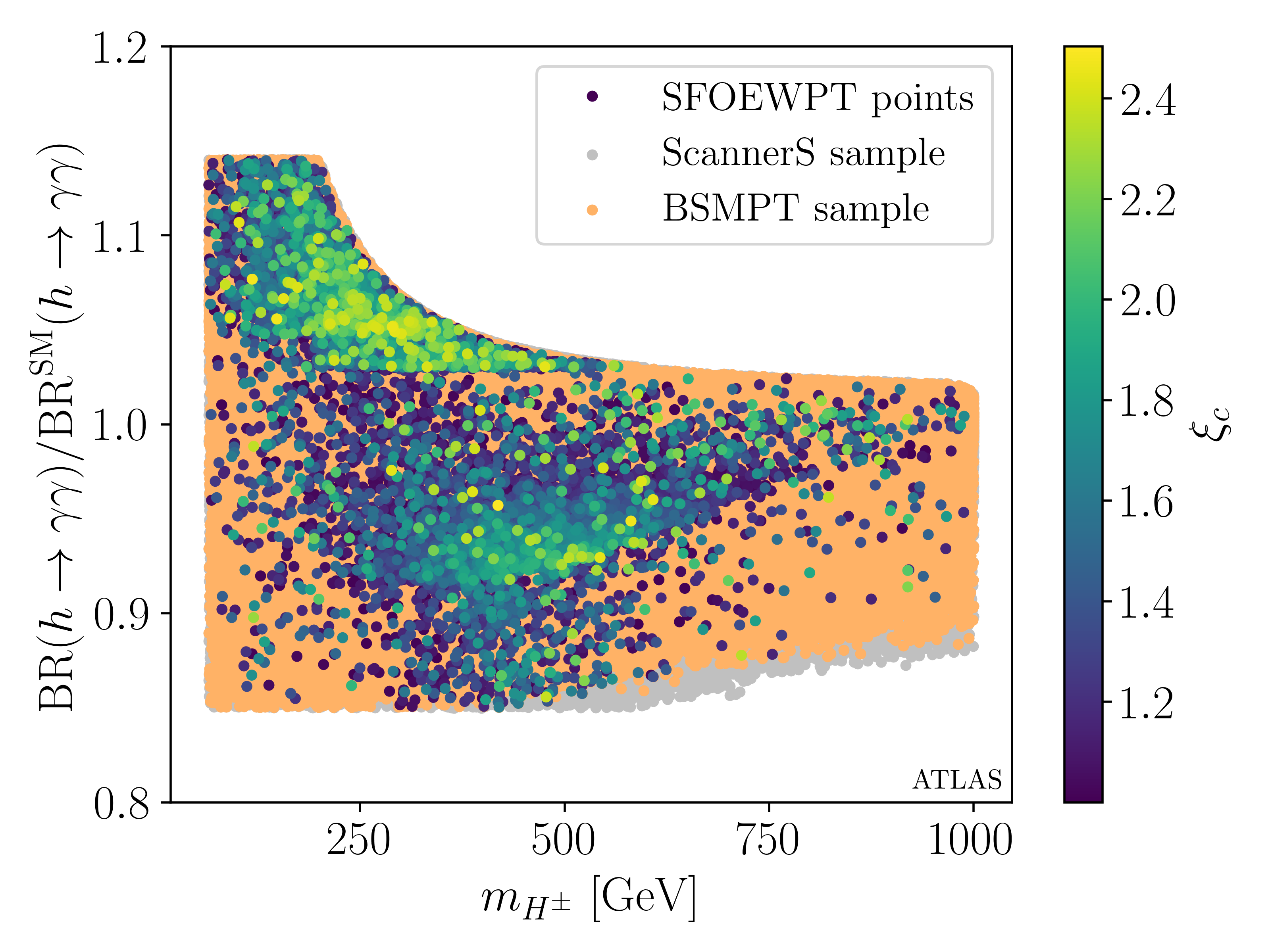}
\includegraphics[width=0.46\textwidth]{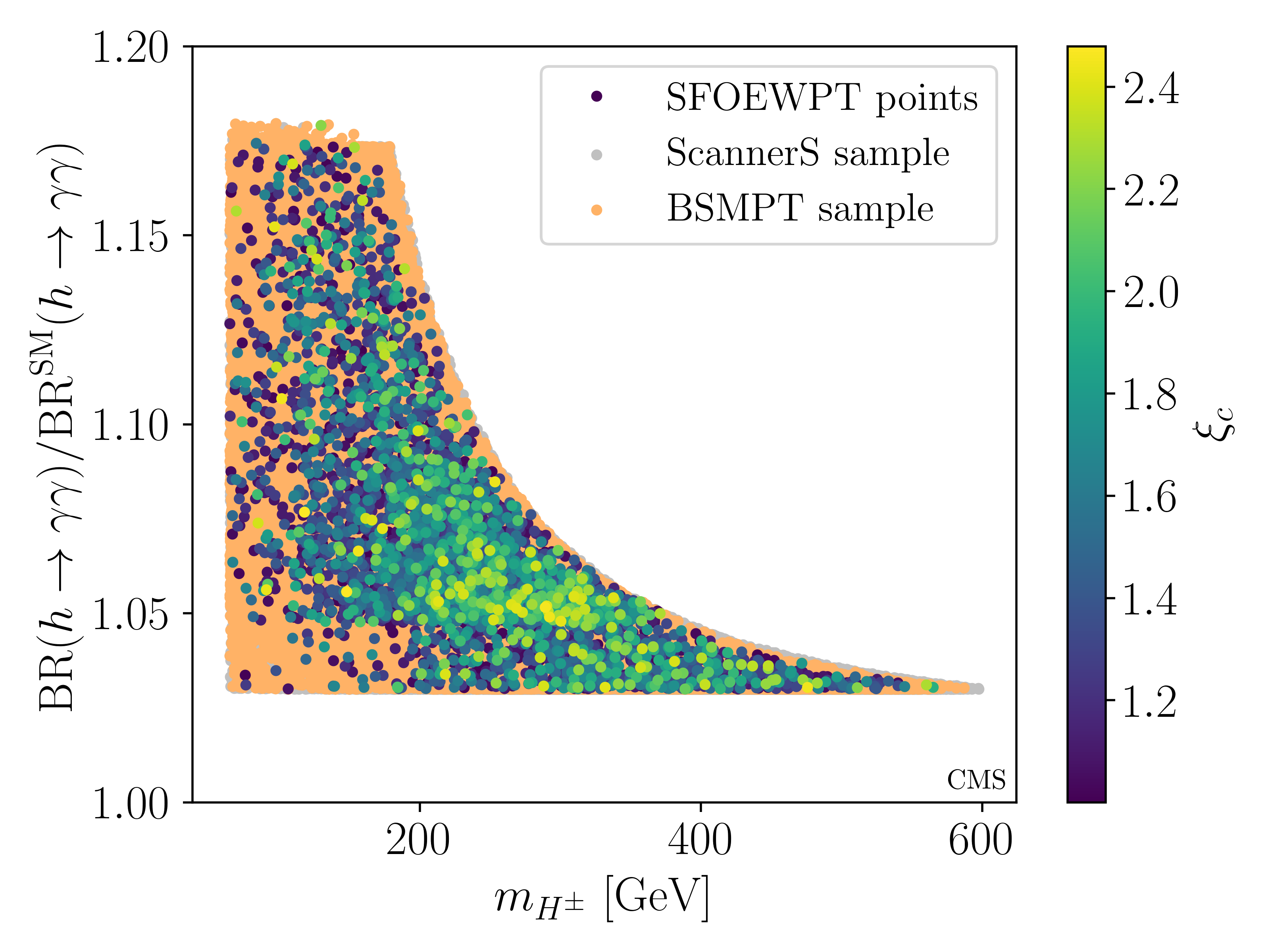}
	\caption{Branching ratio of the SM-like
  Higgs $h$ into a photon pair normalised to the SM value versus
  $m_{H^\pm}$ applying the ATLAS limit (left) and the CMS limit
  (right) on $\mu_{\gamma\gamma}$. Gray points: {\tt ScannerS} sample;
orange points: additionally {\tt BSMPT} constraints imposed; coloured
points: also $\xi_c>1$ fulfilled. Colour code: value of $\xi_c$.} 
	\label{fig:BRhSMgamgamHp}
\end{figure}

In Fig.~\ref{fig:BRhSMgamgamHp} we show the branching ratio of the
SM-like Higgs into two photons normalised to the branching ratio of
the SM Higgs boson as a function of the dark charged Higgs
mass. Neglecting subdominant electroweak corrections, 
the production cross section of the SM-like Higgs,
$\sigma_{\text{prod}} (h)$, is not changed with
respect to that of the SM Higgs boson\footnote{The QCD corrections to
  the production cross section are the same in both models.} so that
the ratio of the branching ratios directly corresponds to the signal rate
$\mu_{\gamma\gamma}$,
\beq
\mu_{\gamma\gamma} \equiv \frac{\sigma_{\text{prod}} (h) \times
  \mbox{BR}(h\to \gamma\gamma)}{\sigma_{\text{prod}}^{\text{SM}} (h)
  \times \mbox{BR}^{\text{SM}} (h\to \gamma\gamma)} \equiv 
\frac{\mbox{BR}(h\to \gamma\gamma)}{\mbox{BR}^{\text{SM}} (h\to
  \gamma\gamma)} \;. 
\eeq
In the left plot we applied the ATLAS limit derived on 
$\mu_{\gamma\gamma}$ \cite{ATLAS:2018hxb} which is given
by
\beq  
\mu_{\gamma\gamma} = 0.99^{+0.15}_{-0.14} \;,  
\eeq  
in the right plot we applied the CMS limit \cite{CMS:2021kom} of
\beq 
\mu_{\gamma\gamma} = 1.12^{\pm 0.09} \;.
\eeq 

The grey points are those that are obtained after applying the {\tt
  ScannerS} constraints described above. The orange points
additionally fulfil the {\tt BSMPT} constraints. In particular {\tt
  BSMPT} checks whether the global minimum at NLO coincides with the
electroweak vacuum. The additional constraints from {\tt
  BSMPT} barely further reduce the {\tt ScannerS} sample.  
The coloured points are those that additionally have
a strong first-order phase transition. The colour code denotes the
strength of the phase transition. We see that in our model we can
reach $\xi_c$ values for the still allowed parameter points that go up
to 2.48. \s

As can be inferred from the plot, the maximum possible branching ratio
values increase towards smaller charged Higgs boson masses. Both the value of
$m_{H^\pm}$ and the $hH^+H^-$ coupling are governed by
$\lambda_3$. The value of the branching ratio into $\gamma\gamma$
increases with negative $\lambda_3$ and the charged Higgs masses
decreases, explaining the behaviour in the plot, {\it cf.}~also
\cite{Azevedo:2018fmj} for a detailed discussion. The parameter space
of our model is constrained by the experimental limits on the photonic
rate, the CMS limit allows for somewhat larger, the ATLAS limit for
smaller $\mu_{\gamma\gamma}$. Note, however, that the allowed points
are cut below the maximally allowed value by CMS of
$\mu_{\gamma\gamma}=1.21$. The upper bound actually results from the
combination of the bounded-from-below and unitarity bounds that
restrict the allowed values of the coupling $\lambda_3$. 
The plots show that a future increased precision in $\mu_{\gamma\gamma}$ can cut the
parameter space on the charged Higgs mass substantially. As can be
inferred from the left plot, the charged Higgs 
mass range starts being cut from $\mu_{\gamma\gamma}$ values above about
1.02 on. In the following
plots, we use {\tt ScannerS} samples that include the more
recent limit on $\mu_{\gamma\gamma}$ which is given by CMS. It reduces
the upper bound on the allowed charged Higgs mass to
597~GeV. The inclusion of the {\tt BSMPT} constraints
reduce it further to 587~GeV, and the requirement of an SFOEWPT to
565~GeV finally. The reduction in $m_{H^\pm}$ in turn also reduces the range of
allowed dark neutral masses as we will see. \s

As for the
parameter points with an SFOEWPT, they are distributed nearly all
over the still allowed parameter space. The demand of an SFOEWPT
hence does not significantly constrain our model with respect to the Higgs
data. While the SFOEWPT limit on $m_{H^\pm}$ is somewhat below the
{\tt BSMPT} limit, a dedicated parameter scan might also provide
SFOEWPT values with larger charged Higgs mass values. 
Vice versa the Higgs rate measurements in photonic final states
do not constrain baryogenesis scenarios of `CP in the Dark'. \s

\begin{figure}[h!]
	\centering
	\includegraphics[width=0.47\textwidth]{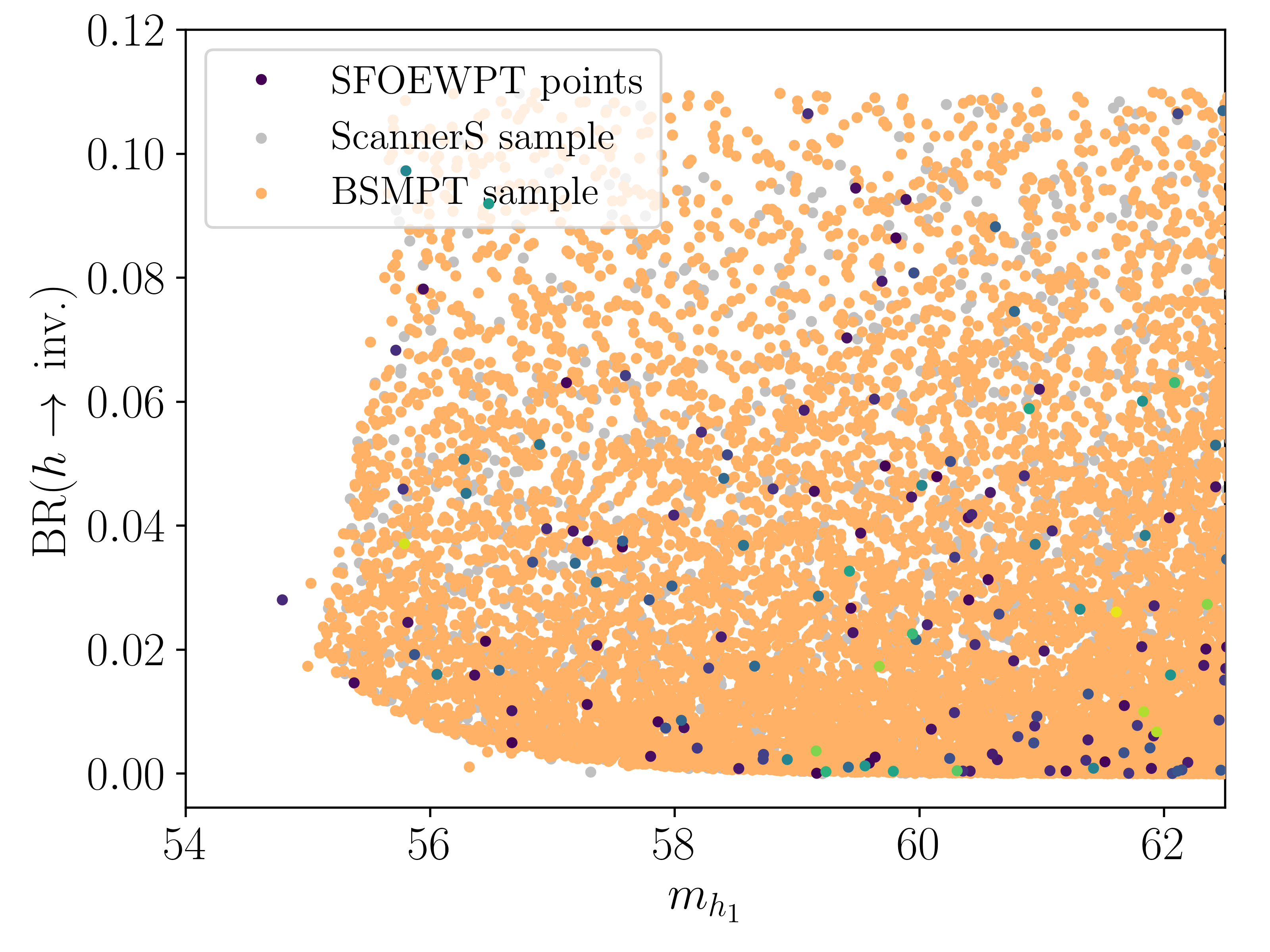}
\includegraphics[width=0.47\textwidth]{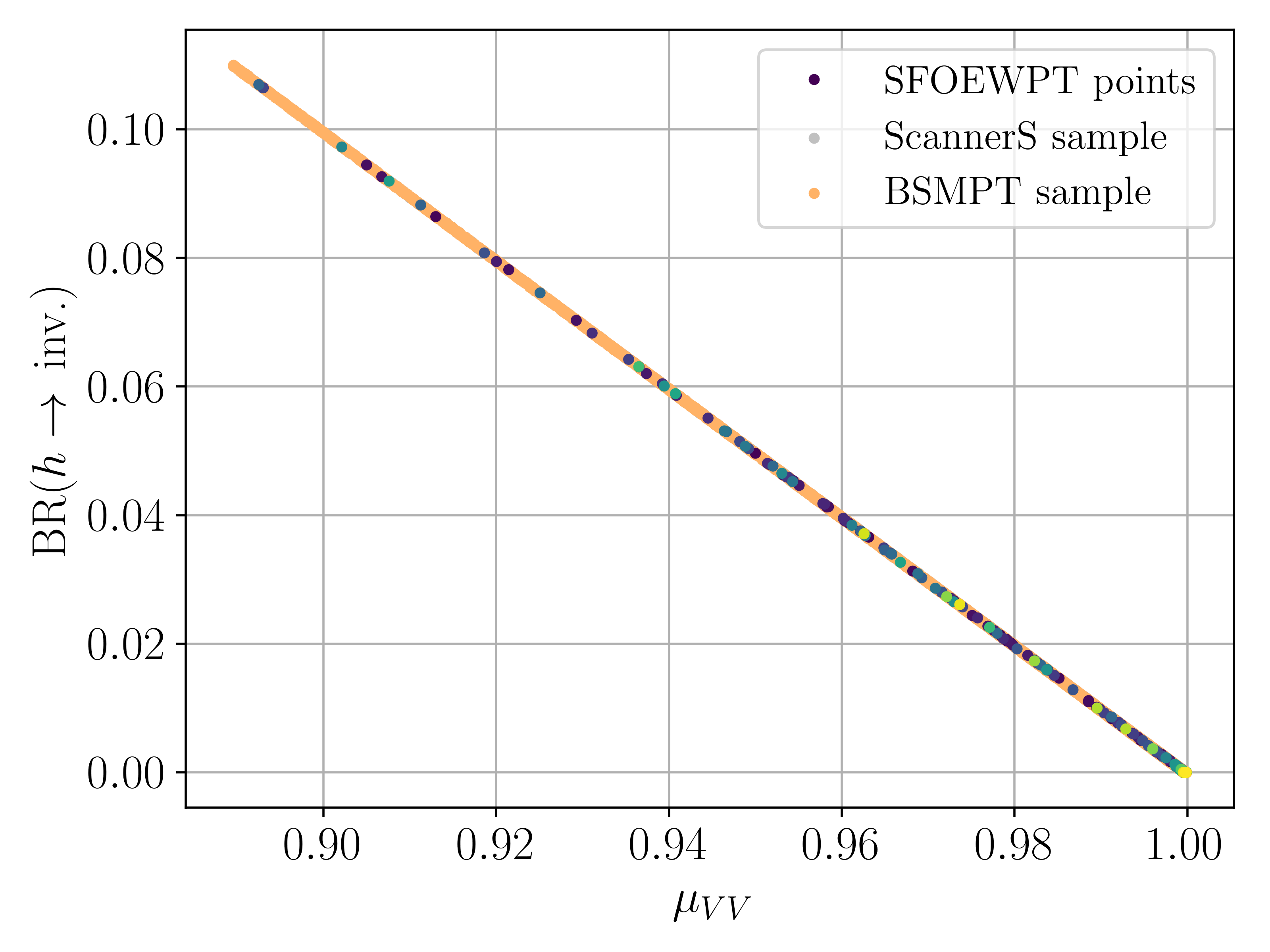}
	\caption{Branching ratio of the SM-like Higgs $h$ into dark
          particles versus $m_{h_1}$ (left) and versus $\mu_{VV}$ (right). The colour code is the same as in \figref{fig:BRhSMgamgamHp}.} 
	\label{fig:BRhSMinvmH1}
\end{figure}

In Fig.~\ref{fig:BRhSMinvmH1} (left) we display the branching ratios of the
SM-like Higgs boson $h$ into invisible particles versus the DM mass
$m_{h_1}$ and in Fig.~\ref{fig:BRhSMinvmH1} (right) versus the gauge
boson signal strength 
$\mu_{VV}$ ($V=Z,W$). 
As mentioned above, we have applied here and in
all other plots presented in the numerical analysis the additional cut on
$\text{BR}(h\rightarrow\,\text{inv.})<\num{0.11}$ following the latest
results of~\cite{ATLAS:2020kdi}. 
We find viable parameter points down to DM masses of about $m_{h_1}=
54.8$~GeV. Below this value it becomes increasingly difficult to find
parameter points that comply with all considered constraints. The
parameter points compatible with an SFOEWPT are scattered across the
still allowed {\tt ScannerS} region. Therefore, future improved
measurements of $h\rightarrow \text{inv.}$ 
are able to test the parameter space of `CP in the Dark' but they will
not give us additional information on the strength of the phase
transition itself. Above $m_{h_1}=62.5$~GeV (not shown in the plot)
the branching ratio of course drops to zero as the corresponding decay
is kinematically closed. \s

The results for $\text{BR}(h\rightarrow\,\text{inv.})$ versus
$\mu_{VV}$ in \cref{fig:BRhSMinvmH1} (right) look similar to those found
in~\cite{Engeln:2020fld} for the fully dark phase (FDP) of the
N2HDM which is very similar to our model. Since all tree-level $h$
couplings are SM-like the invisible branching ratio strongly
correlates with $\mu_{VV}$. It decreases for increasing
$\mu_{VV}$ until $\text{BR}(h\rightarrow\,\text{inv.})=0$ when
$\mu_{VV}=1$. This is expected as for $\mu_{VV}\rightarrow 1$, the
SM-like Higgs branching ratios converge to their SM values with no decays into
invisible particles being allowed. Future precise measurements of the
$\mu_{VV}$ rates will hence constrain the invisible branching ratios
and thereby the parameter space of the model, but again not give further
insights on the strength of the EWPT as can be inferred from the
distribution of the coloured points. 

\FloatBarrier
\subsection{Mass Parameter Distributions for an
  SFOEWPT}\label{subsec:SFOEWPTresults}
\begin{figure}[b!]
	\centering
	\includegraphics[width=0.5\textwidth]{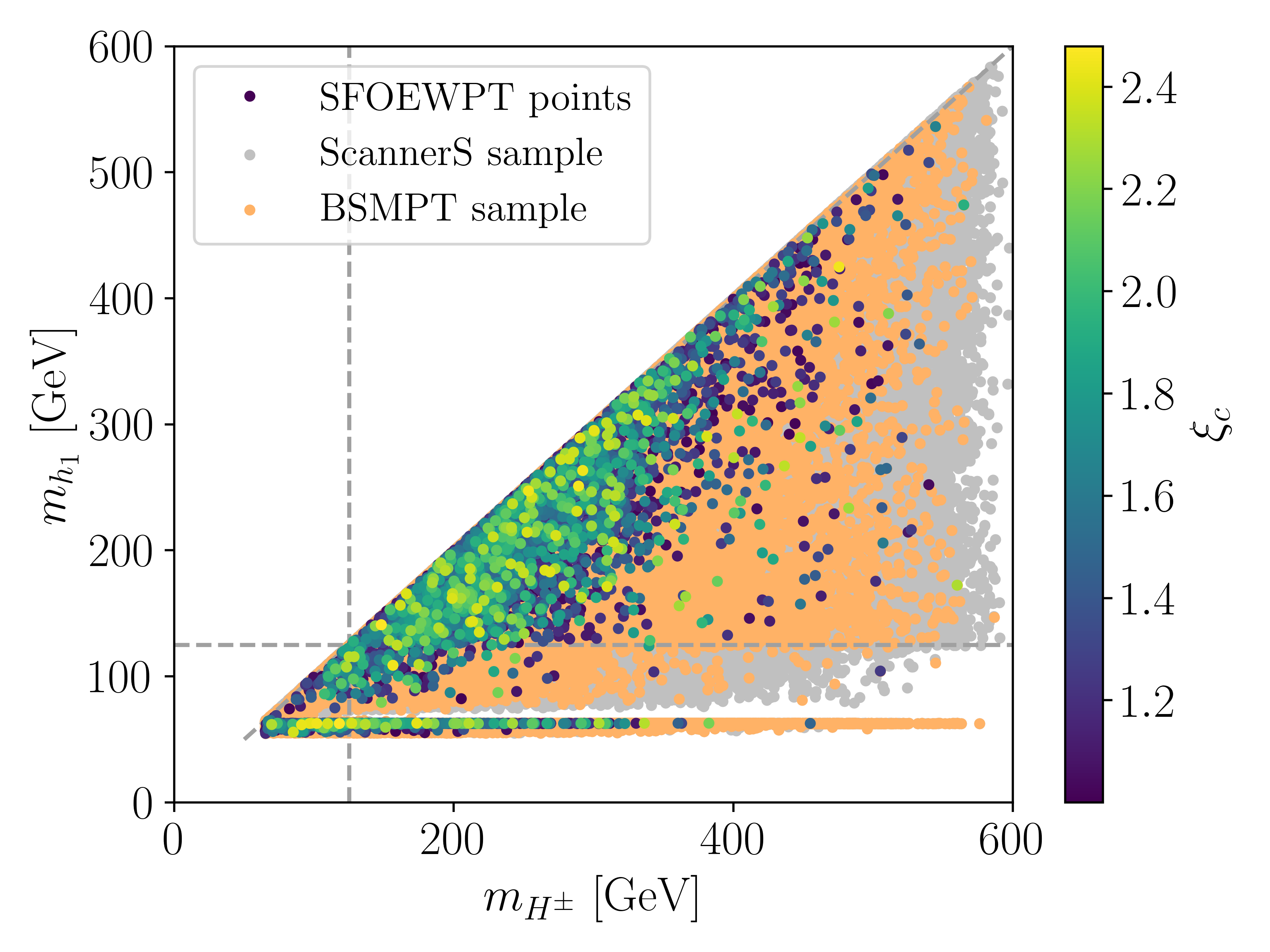}
	\caption{Scatter
  plot in the $m_{h_1}$ and $m_{H^\pm}$ mass plane. The dashed lines are there to guide the eye and denote the $m_{h_1}$ and $m_{H^\pm}$ values at 125~GeV and
  the points where their masses are equal. Colour code as in \figref{fig:BRhSMgamgamHp}.} 
	\label{fig:SFOEWPTscatH1Hpm}
\end{figure}
Figure~\ref{fig:SFOEWPTscatH1Hpm} shows for the parameter points of our
scan the lightest neutral dark scalar mass $m_{h_1}$ versus the dark
charged scalar mass $m_{H^\pm}$. The colour code is the same as in the
previous figures. 
The constraint on the charged Higgs mass values from
$\mu_{\gamma\gamma}$ also constrains the allowed $m_{h_1}$ values
which can go up to 584~GeV in the {\tt ScannerS} sample, to 568~GeV
after inclusion of the {\tt BSMPT} constraints, and reaches a maximum
value of 536~GeV for points providing an SFOEWPT. Depending
on the future restriction on $\mu_{\gamma\gamma}$ the charged Higgs
mass will be less or more constrained with
immediate consequences for the allowed range of $m_{h_1}$. 
The points with $\xi_c \ge 1$ cluster towards smaller mass values. We
still find SFOEWPT points for larger masses, however. A dedicated scan
in this mass region may increase their density. So again, the
requirement of an SFOEWPT does not significantly  
constrain the parameter space nor do the Higgs constraints
further restrict the points leading to $\xi_c$ values above 1. 
We note that the distribution structure of the points stems from the fact
that we performed a dedicated scan in the $m_{h_1}$ mass region below 125/2~GeV
resulting in the horizontally distributed points in the region below 62.5~GeV.
\s
 
\begin{figure}[h!]
	\centering
	\includegraphics[width=0.48\textwidth]{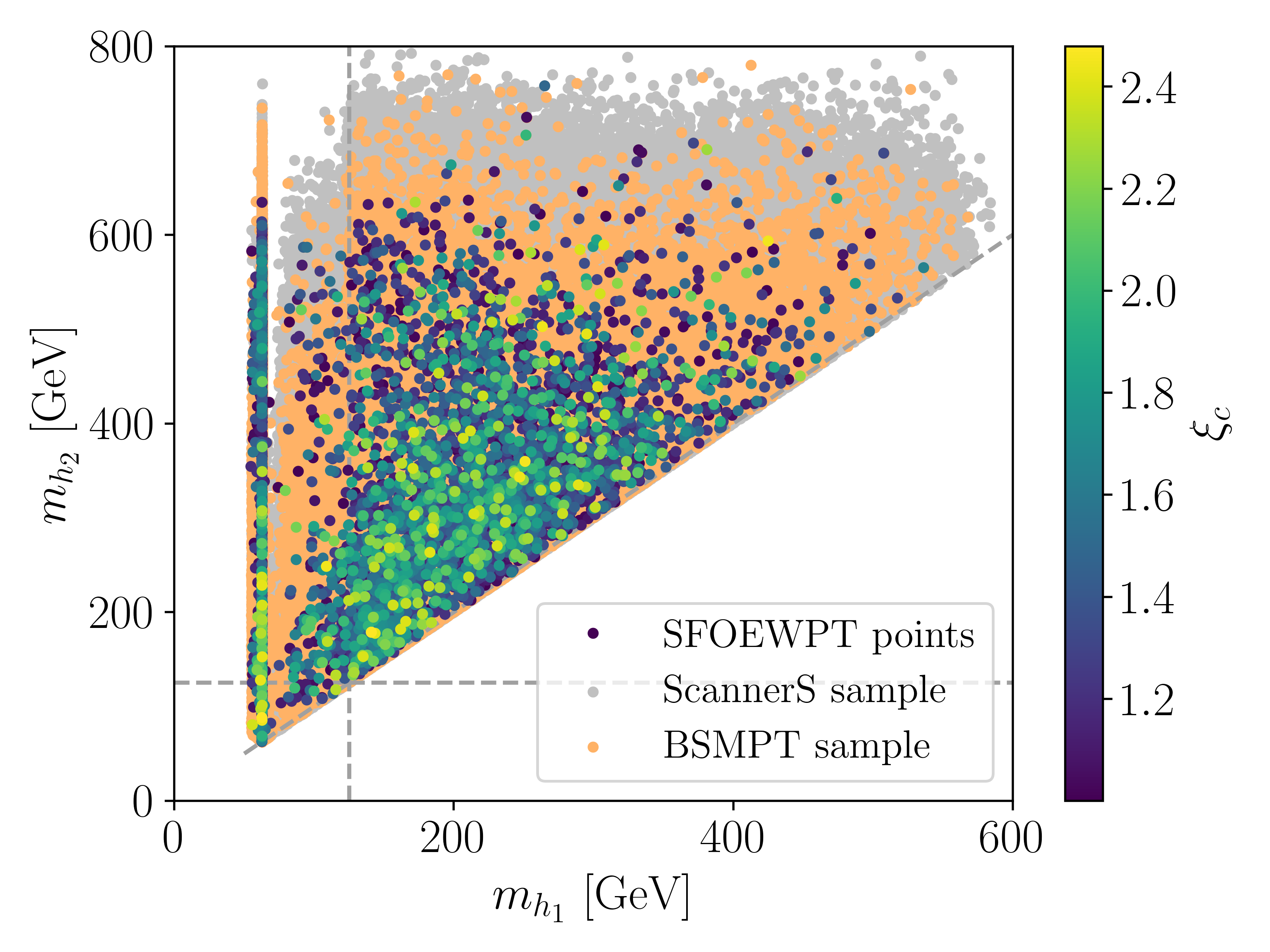}
\includegraphics[width=0.48\textwidth]{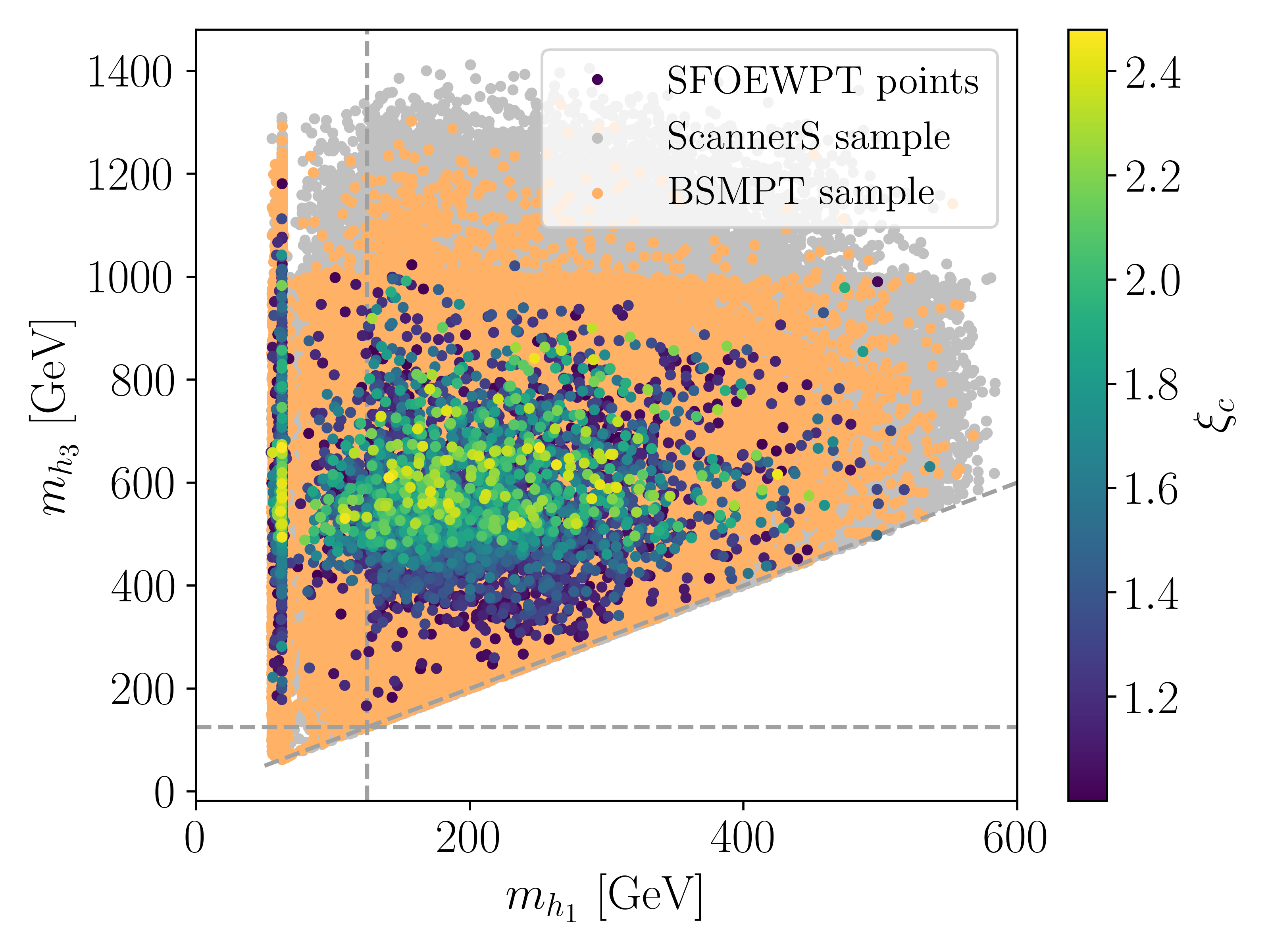}
	\caption{Scatter
  plot $m_{h_1}$ versus $m_{h_2}$ (left) and $m_{h_1}$ versus
  $m_{h_3}$ (right). The dashed lines are there to guide the eye and
  denote neutral dark mass values at 125~GeV and
  the points where their masses in the respective plane are equal. The
  colours denote the same constraints as in 
  Fig.~\ref{fig:BRhSMgamgamHp}.} 
	\label{fig:massdifference}
\end{figure}

In \figref{fig:massdifference} we display the distribution of our
parameter point sample in the neutral DM mass planes, namely $m_{h_1}$
versus $m_{h_2}$ (left) and $m_{h_1}$ versus $m_{h_3}$ (right). Again
the restricted $m_{H^\pm}$ range is reflected in the allowed upper
values of the dark neutral masses. In the $m_{h_1}-m_{h_3}$ plane we
see a tendency of SFOEWPT points to cluster towards smaller mass
values. Still we
have also points for larger values in the allowed BSMPT sample. The requirement
of an SFOEWPT does not allow us to read off strict bounds on the mass values. \s

\FloatBarrier
\subsection{Analysis of the VEV
  Configurations}\label{subsec:VEVEVOresults}
In all our allowed parameter samples we find that the charge-breaking
VEV is zero as required for the photon to remain massless. As for the
other VEVs, at non-zero temperature we find two VEV patterns: In one, the SM-like VEV is
non-zero while the remaining DM VEVs are negligibly small. This is the 
case for almost all allowed parameters sets. The other case is given
by a very small fraction of allowed parameter points. Here we find VEV
configurations where also the dark VEVs develop non-zero values. \s

\begin{figure}[h!]
	\centering
	\includegraphics[width=0.47\textwidth]{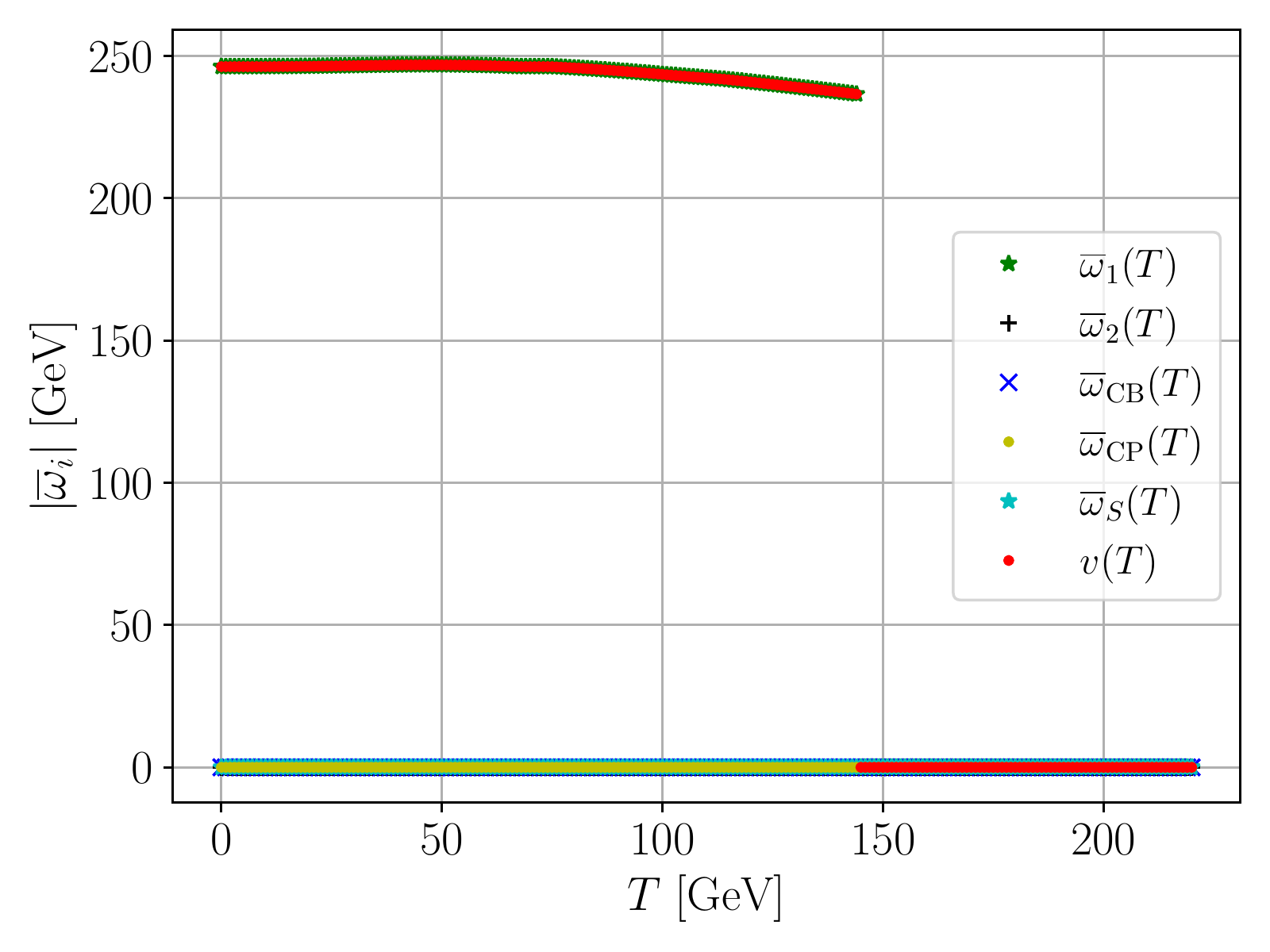}
	\includegraphics[width=0.47\textwidth]{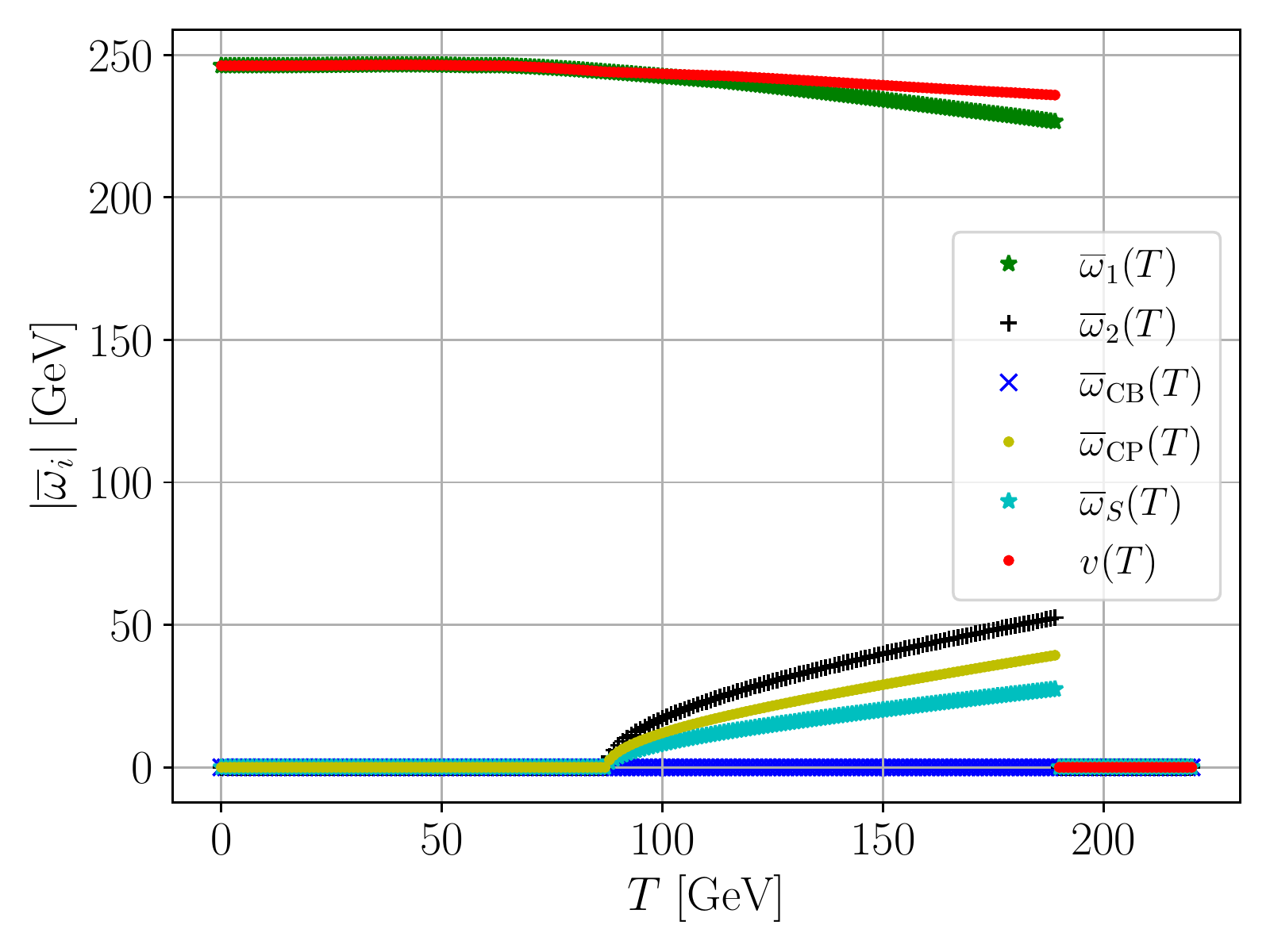}
	\caption{Evolution of the VEVs as a function of $T$ for a
          point where only $|\overline{\omega}_1| \ne 0$ (left) and
          for a point where all except the charge-breaking VEV are
          non-zero at $T_c$. The temperature-dependent electroweak VEV
          $v(T)$ is defined in \eqref{eq:EWVEV} and shown in red. Both
          benchmark points are given in \appen{app:sec:points}. 
	}
	\label{fig:vevevo}
\end{figure}
In \figref{fig:vevevo} we illustrate the evolution of all five VEVs as a
function of the temperature for two sample points, one for each of the two
categories. The sample points are given in \appen{app:sec:points}. In
red, we display the temperature-dependent electroweak VEV $v(T)$, that 
is calculated taking into account the $SU(2)_L$-VEVs, see \eqref{eq:EWVEV}. 
Both points displayed in \figref{fig:vevevo} show a 
discontinuity in $v_c(T)$ at $T=T_c$ that is large enough to be
classified as SFOEWPT. Actually, we have $\xi_c= 1.64$ for the left and
$\xi_c= 1.24$ for the right scenario. Only for the scenario depicted in
the right plot, however, also dark VEVs participate in the SFOEWPT in addition to 
$|\overline{\omega}_1|$ which is non-zero for all $T< T_c$. 
The development of a non-zero CP-violating VEV $|\overline{\omega}_\text{CP}|$
(which remains non-zero down to $T=88$~GeV~$< T_c$ and is zero at zero
temperature) actually
corresponds to the generation of \textit{spontaneous} CP violation. \s

\begin{figure}[ht!]
	\centering
	\includegraphics[width=0.6\textwidth]{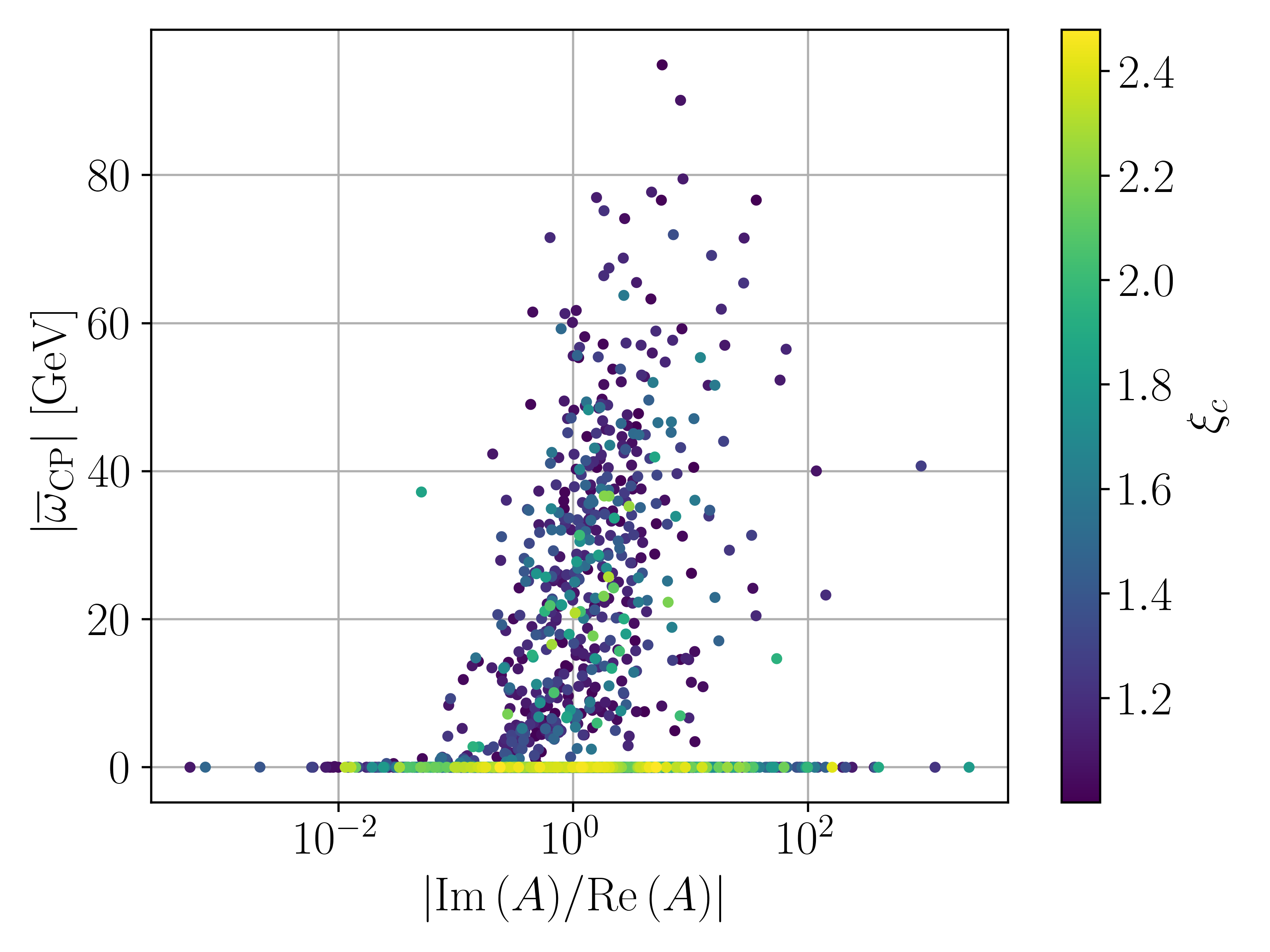}
	\caption{The CP-violating VEV $\overline{\omega}_\text{CP}$
          versus $|\Im (A)/\Re (A)|$ for all SFOEWPT points. The colour code
          indicates the strength $\xi_c>1$ of the SFOEWPT.} 
	\label{fig:vevCPV}
\end{figure}

In Fig.~\ref{fig:vevCPV} we show the absolute value of the CP-violating VEV
$\overline{\omega}_\text{CP}$ as a function of the absolute value of
the ratio $\Im (A)$ over $\Re (A)$ for all allowed SFOEWPT points of our scan.
As discussed in \sect{sec:Model}, a non-zero imaginary part of the
trilinear coupling $A$ induces explicit CP violation. At
$T=0$~GeV, CP violation can only be generated explicitly, as
$\overline{\omega}_{\text{CP}}\big|_{T=\SI{0}{GeV}}\equiv v_\text{CP} = 0$. We find
in total 564 points for which $|\overline{\omega}_\text{CP}| \ne 0$
(more specifically, $|\overline{\omega}_\text{CP}| > 0.001$~GeV) 
at finite temperature and which hence break CP \textit{spontaneously}
at $T>0$. Additionally, these points 
develop a non-vanishing singlet VEV $\overline{\omega}_S$ at $T \ne 0$. This
means that also the $\nZ_2$-symmetry is spontaneously
broken. At finite temperature, the dark charge therefore is not
conserved, and particles that are dark at zero temperature can now mix
with particles from the first doublet. This is very interesting 
as it provides a promising portal for the transfer of
non-standard CP violation to the SM-like Higgs couplings to fermions at 
finite temperature. This is in addition to an SFOEWPT another
necessary ingredient for an EWBG scenario that is able to explain
today's observed BAU. We finally note, that the plot does not show a
clear correlation between the size of $|\overline{\omega}_{\text{CP}}|$ and $\Im
(A) \ne 0$. However, we see that $|\overline{\omega}_{\text{CP}}| >
0$ only for $\Im (A) \ne 0$. From the plot we cannot deduce a correlation
between the size of $\xi_c$ and $|\overline{\omega}_{\text{CP}}|$: For the
strength of the phase transition, overall the participation of
additional Higgs bosons and their involved mass values is decisive. It
is not important which kind of VEV contributes to $v_c(T)$.

\FloatBarrier
\subsection{Dark Matter Observables}\label{subsec:DMobservables}
\begin{figure}[ht!]
    \centering 
    \includegraphics[scale=0.6]{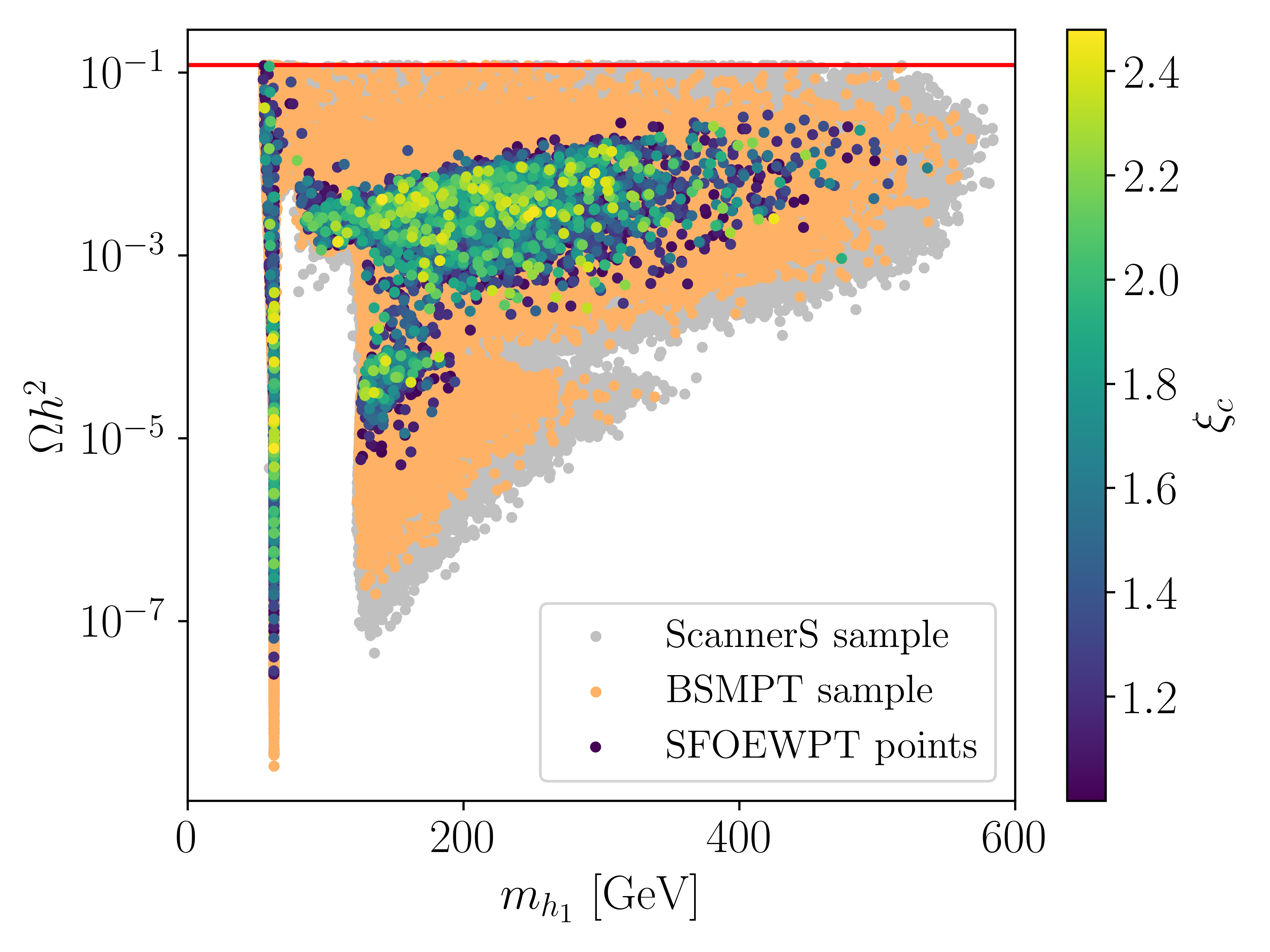} 
    \caption[Relic density versus dark matter mass $m_{h_1}$ for the 
    benchmark parameter sample]{Relic density versus DM mass $m_{h_1}$
      for the benchmark parameter sample, with the colour code 
      introduced in \figref{fig:BRhSMgamgamHp}. The experimentally 
      measured relic density $\Omega_\text{obs} h^2 =
      \num{0.1200\pm0.0012}$~\cite{Aghanim:2018eyx} is shown in 
      red.}  
    \label{fig:relicdensity} 
\end{figure}

In \figref{fig:relicdensity} we show our benchmark point sample in the
plane spanned by the relic density $\Omega h^2$, calculated via
\texttt{ScannerS} through the link with {\tt MicrOMEGAs}, and the mass
of the DM candidate, $m_{h_1}$. The experimentally measured
relic density $\Omega_\text{obs} h^2 =
\num{0.1200\pm0.0012}$~\cite{Aghanim:2018eyx} is shown in red. The
colour code is the same as in \figref{fig:BRhSMgamgamHp}. While we find
{\tt ScannerS} sample points that lie within the $1\sigma$ error bands for the measured
relic density, the SFOEWPT points are all underabundant.\footnote{Due
to the logarithmic scale this cannot be inferred from the plot by
eye.} Parameter samples with masses around $m_{h_1}/2$ can be less
underabundant than scenarios with heavier DM particles. 
The underabundance is not problematic. It simply means that we need another
DM component to make up for the total of the relic density. We can
hence state that the requirement of an SFOEWPT in `CP in the Dark' is
compatible with the measured DM relic density. \s

In order to investigate the impact of the measurements of the direct
detection spin-independent (SI) nucleon DM cross section $\sigma$ we
first compute the effective cross section $f_{\chi \chi}\cdot\sigma$
for our model,
\begin{align}
  \sigma\cdot f_{\chi\chi} \equiv \sigma\cdot \frac{\Omega_\text{prod}h^2}{\Omega_\text{obs}h^2}\,.\label{eq:EffSIDDCS}
\end{align}
\begin{figure}[hb!]
    \centering \includegraphics[scale=0.6]{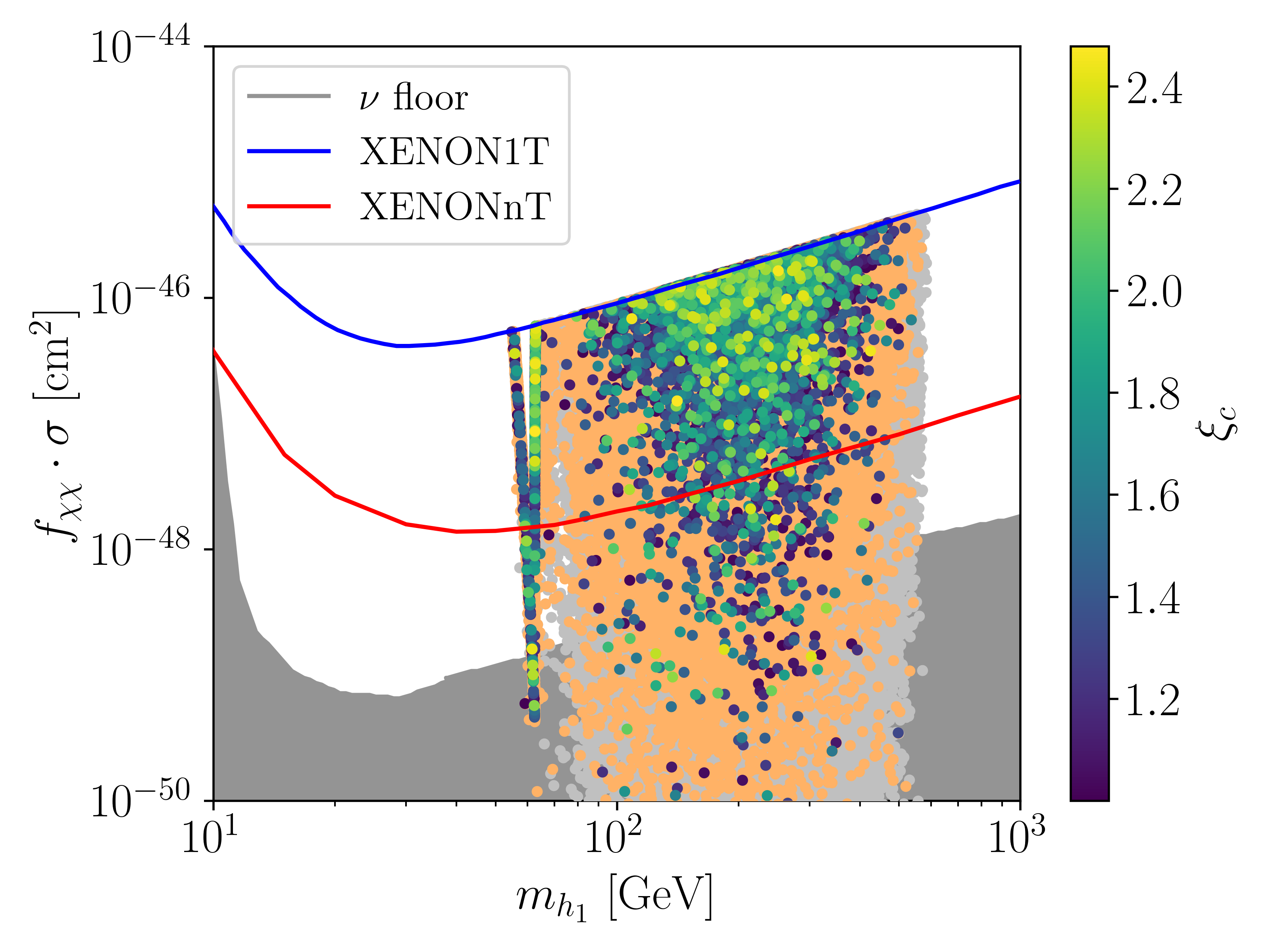} 
    \caption[Effective direct detection SI nucleon DM cross section of
    the benchmark point sample versus $m_{h_1}$]{Effective direct
      detection SI nucleon DM cross section of the benchmark point
      sample versus $m_{h_1}$, with the colour code defined as in
      \figref{fig:BRhSMgamgamHp}. The experimental results have
      been obtained by using~\cite{dmtools}. The {\tt XENON1T}
      exclusion limit~\cite{Aprile:2018dbl} is shown in blue. In red,
      we show the projected sensitivity of the {\tt XENONnT}
      experiment~\cite{Aprile:2020vtw}. The experimental limit for the
      neutrino background has been taken from~\cite{Billard:2013qya}.}  
    \label{fig:DDnSIlimits} 
\end{figure}
The rescaling factor $f_{\chi\chi}$ considers the fact that
in our model, depending on the parameter point, the relic density
$\Omega_\text{prod}h^2$ can be underabundant, which has to be taken
into account when comparing with the measured value of $\sigma$, {\it cf.}~also
\cite{Engeln:2020fld,Glaus:2020ihj}. The numerical values for the
produced relic density $\Omega_\text{prod}h^2$ in our model are obtained using {\tt
  MicrOMEGAs}. In \figref{fig:DDnSIlimits} we display the effective direct
detection SI nucleon DM cross section $f_{\chi \chi}\cdot\sigma$ of
the benchmark point sample versus $m_{h_1}$. 
As already required by the constraints in {\tt ScannerS} (linked to
{\tt MicrOMEGAs}), all points
lie below the {\tt XENON1T} exclusion limit, which is displayed in
blue. The majority of the SFOEWPT points is found to be above the neutrino
floor (dark grey shaded area). In addition, most SFOEWPT points are also above the
expected sensitivity of the {\tt XENONnT} experiment (red). This means
that future DM direct detection
experiments will allow us to test a large fraction of the parameter
space of `CP in the Dark' that is compatible with an
SFOEWPT. 

\FloatBarrier
\section{Conclusion}\label{sec:Conclusion}
In this paper, we investigated the possibility of an SFOEWPT within
the framework of the model `CP in the Dark'. Its extended N2HDM-like
scalar sector provides a dark sector that is stabilised at zero
temperature through only one $\nZ_2$ symmetry and thereby provides a DM
candidate. We discussed the treatment of finite pieces of our
renormalisation scheme and the necessary adjustments for this model in
contrast to past work to be able to renormalise the one-loop mixing
angles and masses to their leading-order values. This allows us to
efficiently perform parameter scans taking into account the relevant
theoretical and experimental constraints to obtain viable parameter
sets. The new {\tt BSMPT} version {\tt v2.3} including the
  implementation of `CP in the Dark' has been made publicly available
  under \url{https://github.com/phbasler/BSMPT}.
\s 
 
Our results show that `CP in the Dark' proves itself to be a highly promising candidate to
explain the BAU in an EWBG context, as in addition to explicit 
CP violation in the dark sector, it also provides spontaneous CP violation at finite
temperature. In combination with the also spontaneously broken
$\nZ_2$ symmetry at non-zero temperature non-standard CP violation
can be transferred to the couplings of the SM-like Higgs boson to
fermions. This may allow for a large enough CP violation to generate the
BAU observed by experiment, without being in conflict with the EDM constraints. \s

Viable SFOEWPT points are distributed across almost the
whole allowed dark mass ranges of the model. While the SM-like Higgs
rates will allow us to constrain the parameter space of the model, the
SFOEWPT points do not impose further significant constraints. On the other hand,
SFOEWPT points are found to be in the reach of precise measurements of
invisible decays of the SM-like Higgs boson and of the Higgs rates
into SM particles at the LHC. Our SFOEWPT points comply with the measured
relic density. We found that a large fraction of the parameter space
and SFOEWPT points 
of the model will be testable at future DM direct detection
experiments. \s 

Having demonstrated in this paper that all prerequisites for
BAU are fulfilled in the model, the next natural steps to be taken in future work is the
implementation of the computation of the amount of baryogenesis in
{\tt BSMPT} in order to investigate if the model can indeed provide
the correct amount of BAU. If this is the case, subsequently LHC and
DM observables are to be identified that may serve as smoking gun
signatures for parameter scenarios compatible with BAU in `CP in the
Dark'. 

\subsection*{Acknowledgments}
The research of MM was supported by the Deutsche
Forschungsgemeinschaft (DFG, German Research Foundation) under grant
396021762 - TRR 257.  JM acknowledges support by the BMBF-Project 05H18VKCC1.
We thank Philipp Basler, Duarte Azevedo, G\"unter Quast, Jonas Wittbrodt and Michael
Spira for fruitful discussions.  


\appendix
\section{Benchmark Points}\label{app:sec:points}
The input values of the benchmark points discussed in
Subsec.~\ref{subsec:VEVEVOresults} are given in
Tab.~\ref{tab:benchpoints}. The dark mass values, critical
temperature, critical VEV, $\xi_c$ and the individual VEVs at $T_c$ are given
in Tab.~\ref{tab:bench2}. Note that we have $\lambda_1\simeq\num{0.258}$,
  $m_{11}^2\simeq\num{-7824}$~GeV$^2$ for both points. The parameter
  $\lambda_1$ is fixed through $m_h^2 = \lambda_1 v_1^2$ and the value for $m_{11}^2$ follows from the minimisation condition.

\begin{table}[h!]
\begin{center}
    \begin{tabular}[]{l c c c c c c}\toprule
        \text{point}&$m_{22}^2\, [\si{GeV}^2]$&$m_S^2\,
[\si{GeV}^2]$&$\text{Re}\,(A)\, [\si{GeV}]$&$\text{Im}\,(A)\, [\si{GeV}]$
      &$ \lambda_1$\\\midrule
        Fig.~5(a)&$\num{96703.414}$&$\num{32442.949}$&$\num{159.627}$&$\num{-325.391}$
      & $\num{3.532}$ \\
        Fig.~5(b)&$\num{65258.809}$&$\num{36279.847}$&$\num{279.502}$&$\num{-326.645}$
      & $\num{3.660}$ \\ \hline \hline
\text{point} & $\lambda_3$ &$\lambda_4$ & $\lambda_5$ & $\lambda_6$
                 & $\lambda_7$ & $\lambda_8$\\ \midrule
        Fig.~5(a)&$\num{-0.796}$&$\num{0.787}$&$\num{-0.055}$&$\num{10.446}$&$\num{7.596}$&$\num{4.683}$\\
        Fig.~5(b)&$\num{-0.821}$&$\num{0.220}$&$\num{-0.371}$&$\num{4.715}$&$\num{7.760}$&$\num{14.781}$\\\bottomrule
    \end{tabular}
\caption{Input parameters of the two benchmark points of
  Fig.~\ref{fig:vevevo}. We set $v_1 \equiv v \approx 246.22$~GeV and
  $m_h=125.09$~GeV.\label{tab:benchpoints}} 
\end{center}
\end{table}

\begin{table}[h!]
\begin{center}
    \begin{tabular}[]{l c c c c c c}\toprule
        \text{point}&$m_{H^\pm}$&$m_{h_1}$&$m_{h_2}$&$m_{h_3}$ & $T_c$&$v_c$ 
\\ \midrule
        Fig.~5(a)&$\num{269.386}$&$\num{241.718}$&$\num{308.943}$&$\num{549.265}$ & $\num{144.21}$&$\num{236.53}$ \\
        Fig.~5(b)&$\num{200.940}$&$\num{62.680}$&$\num{218.700}$&$\num{560.206}$ & $\num{189.77}$&$\num{235.85}$\\ \hline \hline 
        \text{point}& $\xi_c$ &$\overline{\omega}_\text{CB}$&$\overline{\omega}_1$&$\overline{\omega}_2$&$\overline{\omega}_\text{CP}$&$\overline{\omega}_S$\\\midrule
        Fig.~5(a)&$\num{1.64}$&$\num{-8.977e-07}$&$\num{236.53}$&$\num{9.093e-07}$&$\num{-3.793e-07}$&$\num{4.604e-07}$\\
        Fig.~5(b)&$\num{1.24}$&$\num{-2.212e-05}$&$\num{226.46}$&$\num{52.72}$&$\num{39.52}$&$\num{-27.58}$\\\bottomrule
    \end{tabular}
\caption{Dark mass values, critical temperature and critical VEV, $\xi_c$, and
  individual VEVs at $T_c$ of the two benchmark points of
  Fig.~\ref{fig:vevevo}. All values, except for $\xi_c$, are given in
  GeV. \label{tab:bench2}} 
\end{center}
\end{table}


\bibliographystyle{h-physrev}
\bibliography{N2HDM_Paper}



\end{document}